\documentclass[a4paper,11pt]{article}

\PassOptionsToPackage{color}{xy}

\usepackage[margin=2.2cm]{geometry}

\usepackage{mathrsfs}
\usepackage{dsfont}
\usepackage{bbm}
\usepackage{palatino}

\usepackage[usenames,dvipsnames]{xcolor}
\usepackage{graphicx, calc}
\graphicspath{{./figures/}}
\usepackage[font=small]{caption}
\usepackage{titlesec}
\usepackage{subcaption}
\usepackage{float}
\usepackage{array}
\usepackage{multirow}
\usepackage{bm}
\usepackage{amsmath}
\usepackage{amssymb}
\usepackage{amsthm}  
\usepackage{stmaryrd}
\usepackage[colorlinks=true,urlcolor=black,linkcolor=Mahogany,citecolor=Mahogany,plainpages=false,pdfpagelabels]{hyperref}

\newcolumntype{X}[1]{>{\centering\let\newline\\\arraybackslash}p{#1}}

\usepackage[backend=bibtex,maxcitenames=4,maxalphanames=100,maxbibnames=100,isbn=false]{biblatex}
\usepackage{tikz}
\usepackage{verbatim}
\usetikzlibrary{shapes.geometric,plotmarks,backgrounds,fit,calc,circuits.ee.IEC}
\usetikzlibrary{decorations.pathreplacing}

\usepackage{caption}
\usepackage{subcaption}
\usepackage{ragged2e}
\DeclareCaptionJustification{justified}{\justifying}
\usepackage{qcircuit}
\def\A#1{\save []="#1" \restore}
\def\qww{\qw & \qw}
\def\mX{\measure{\makebox[.9em][c]{$X$}}}
\def\mZ{\measure{\makebox[.9em][c]{$Z$}}}

\newcommand{\meterb}[1]{*=<1.8em,2.2em>{\xy 0;<0em,-.8em>:
0*{\begingroup
\everymath{\scriptstyle}
\tiny #1 \endgroup},<0em,.3em>*{\xy ="j","j"-<.778em,-.322em>;{"j"+<.778em,.322em> \ellipse ur,_{}},"j"-<0em,-.2em>;p+<.5em,.9em> **\dir{-},"j"+<2.2em,2.2em>*{},"j"-<2.2em,2.2em>*{} \endxy} 
\endxy} \POS ="i","i"+UR;"i"+UL **\dir{-};"i"+DL **\dir{-};"i"+DR **\dir{-};"i"+UR **\dir{-},"i" \qw}

\newcommand{\pone}{\mathcal{Q}_1} 

\makeatletter
\renewcommand{\fnum@figure}{Fig. \thefigure}
\makeatother

\newcommand{\ket}[1]{{\ensuremath{\lvert#1\rangle}}}

\newcommand{\oline}{\overline}
\newcommand{\cl}{\mathcal}

\newlength\myheight
\newlength\mydepth
\settototalheight\myheight{Xygp}
\DeclareMathOperator{\wt}{wt} 
\DeclareMathOperator{\pr}{Pr} 

\DeclareMathOperator{\eqdef}{:=}

\DeclareMathOperator{\cnot}{CNOT}

\newtheorem{definition}{Definition}
\newtheorem{assumption}{Assumption}

\newtheorem{procedure}{Procedure}
\newtheorem*{procedure*}{Procedure}
\newtheorem{lemma}{Lemma}
\newtheorem{theorem}{Theorem}
\newtheorem{remark}{Remark}

\titlespacing*{\paragraph}{\parindent}{1.8ex plus .2ex minus .2ex}{2ex plus .5ex minus .2ex}

\begin{document}
\title{\vspace{-20mm}{Factory-Based Fault-Tolerant Preparation of Quantum Polar Codes Encoding One Logical Qubit}}

\author{Ashutosh Goswami$^1$, \qquad Mehdi Mhalla$^2$, \qquad Valentin Savin$^3$\\[2mm]
    {\small $^1$\,QMATH, Dep. of Mathematical Sciences, Univ. of Copenhagen} \\
    {\small Universitetsparken 5, 2100 Copenhagen, Denmark}\\
    {\small $^2$\,Univ. Grenoble Alpes, CNRS, Grenoble INP, LIG, F-38000 Grenoble, France}\\
    {\small $^3$\,Univ. Grenoble Alpes, CEA-Leti, F-38054 Grenoble, France}\\
    {\small akg@math.ku.dk, mehdi.mhalla@univ-grenoble-alpes.fr, valentin.savin@cea.fr}
}
\date{}
\maketitle


\begin{abstract}
A fault-tolerant way to prepare logical code-states of $\pone$ codes, i.e., quantum polar codes encoding one qubit, has been recently proposed. The fault tolerance therein is guaranteed by an error detection gadget, where if an error is detected during the preparation, one discards entirely the preparation. Due to error detection, the preparation is probabilistic, and its success rate, referred to as the \emph{preparation rate}, decreases rapidly with the code-length, preventing the preparation of code-states of large code-lengths. In this paper, to improve the preparation rate, we consider a factory preparation of $\pone$ code-states, where one attempts to prepare several copies of $\pone$ code-states in parallel. Using an extra scheduling step, we can avoid discarding the preparation entirely, every time an error is detected, hence, achieving an increased preparation rate in turn. We further provide a theoretical method to estimate preparation and logical error rates of $\pone$ codes, prepared using factory preparation, which is shown to tightly fit the Monte-Carlo simulation based numerical results. Therefore, our theoretical method is useful for providing estimates for large code-lengths, where Monte-Carlo simulations are practically not feasible. Our numerical results, for a circuit-level depolarizing noise model, indicate that the preparation rate increases significantly, especially for large code-length $N$. For example, for $N = 256$, it increases from $0.02\%$ to $27\%$ for a practically interesting physical error rate $p = 10^{-3}$. Remarkably, a $\pone$ code with $N = 256$ achieves logical error rates around $10^{-11}$ and $10^{-15}$ for $p = 10^{-3}$ and $p = 3  \times 10^{-4}$, respectively. This corresponds to an improvement of about three orders of magnitude compared to a surface code with similar code-length and minimum distance, thus showing the promise of the proposed scheme for large-scale fault-tolerant quantum computing.





\end{abstract}


\section{Introduction}

 Polar codes are known for their execellent error correction performance in both classical and quantum communication settings~\cite{arikan2009channel, renes2011efficient,renes2014polar,wilde2013polar, dupuis2021polarization}.  In particular, they achieve the symmetric capacity of any binary-input discrete memoryless classical channel, and achieve the symmetric coherent information of any quantum channel. They also come equipped with a fast classical decoding algorithm (log-linear complexity), which can readily be adapted for Pauli channels, using a syndrome based decoding approach~\cite{renes2011efficient, dupuis2021polarization}.


Despite their excellent error correction performance, and nice algebraic and structural properties, polar codes remained largely unexplored for fault-tolerant quantum computation (FTQC). The main reason is due to the high weight of their stabilizer group generators, which prevents fault-tolerant state preparation and error correction from being implemented by repeated syndrome measurements. This contrasts with conventional approaches based on topological, or more generally quantum low-density parity-check (LDPC) codes, allowing the implementation of a fault-tolerant quantum memory through repeated syndrome measurements, with errors being detected by the difference between syndromes measured in consecutive rounds~\cite{kitaev2003fault, bombin2006topological, breuckmann2021quantum}.

Recently, a fault-tolerant procedure to prepare logical states of $\pone$ codes, that is, CSS quantum polar codes encoding one logical qubit, has been proposed in~\cite{goswami2022fault}. Combined with Steane error correction~\cite{steane1997active, steane2002fast}, this preparation procedure provides an alternative and promissing approach for building a fault-tolerant quantum memory. 


\smallskip The preparation in~\cite{goswami2022fault} is measurement-based, where a set of $N$ qubits are first initialized in the Pauli $Z$ basis and then two qubit Pauli measurements are recursively applied on them. To achieve fault tolerance, the preparation is aided by an error detection gadget, which detects errors at each level of recursion. For error detection-aided preparation, it has been explicitly proven that the preparation is fault-tolerant, in the sense that the weight of the error in the prepared state does not exceed the number of components that fail during the procedure. The fault tolerance therein has been further confirmed by numerical simulations, revealing practically interesting pseudo-threshold values for small $\pone$ codes of length $N \leq 64$ qubits, and showing the promise of the proposed approach to fault-tolerant error correction.

\smallskip However, the preparation of $\pone$ code-states in~\cite{goswami2022fault} is probabilistic due to the error detection gadget. If the gadget detects an error at some recursion level, one declares a preparation failure and discards the prepared state. Hence, one may need to restart the preparation from the beginning several times before a $\pone$ code is successfully prepared. The preparation rate, $i.e.$, the rate of successful preparation, decreases rapidly with the code-length and approaches zero as the code-length increases, hence preventing the preparation of large $\pone$ code-states.

\smallskip In this paper, to improve the rate of successful preparation, we consider a factory preparation of $\pone$ code-states, where one attempts to prepare several copies of polar code-states in parallel, using the measurement-based preparation with error detection from~\cite{goswami2022fault}. Taking advantage of the recursive nature of the preparation, we introduce an extra scheduling step at some recursion levels, so that even if errors are detected we may proceed to the next level of recursion. In other words, we may not need to restart the whole procedure from the beginning every time an error is detected. Therefore, the factory preparation may provide better preparation rates compared to the preparation in~\cite{goswami2022fault}.

\smallskip In addition, we conduct a thorough theoretical analysis of the proposed factory-based preparation approach, constituting one of the most significant contributions of the paper. To this end, we define the notions of rough and smooth errors depending on whether or not the error flips one of the measurement outcomes in the measurement-based preparation. Using rough and smooth errors, we provide theoretical estimates of the preparation rate, as well as the probability of $X$ and $Z$ errors on the prepared state. The latter are used to estimate the logical error rates of the $\pone$ codes under Steane error correction, by using density evolution, as in ~\cite[Section V.D]{goswami2022fault}. Our theoretical estimates are further substantiated by Monte-Carlo simulations. For the circuit level depolarizing noise model, we observe that our theoretical estimate of the preparation rate fits well the Monte-Carlo simulation for code-lengths $N = 64$ and $N = 256$. Further, for the code-length $N= 64$, our theoretical estimate of the logical error rate matches the logical error rate obtained based on Monte-Carlo simulation. Therefore, we use our theoretical estimates to obtain logical error rates corresponding to small physical error rates and large code lengths, where Monte-Carlo simulation is not practically feasible.

\smallskip
For the circuit level depolarizing noise model, our numerical results show that the factory preparation improves significantly the preparation rate of $\pone$ code-states compared to~\cite{goswami2022fault}. In particular, for the physical error rate $p = 10^{-3}$, the preparation rate increases from $47\%$ to $70\%$ for a $\pone$ code-state of length $N = 64$ and from $0.02\%$ to $27\%$ for a $\pone$ code of length $N = 256$. The improvement for $N = 256$ is quite significant as the preparation rate of $27\%$ is practically feasible and implies a qubit overhead only by a factor of four. We have further included numerical results on the logic error rates of $\pone$ codes, using Steane error correction that incorporates our factory preparation of $\pone$ code-states. The $\pone$-code of length $N = 256$ achieves a logical error rate of $10^{-11}$ and $10^{-15}$ for  physical error rate of $10^{-3}$ and $3 \times 10^{-4}$, respectively. A comparison with a surface code of similar length and minimum distance is also provided, further reinforcing the promise of polar codes for fault-tolerant quantum computation.
 
\smallskip  The paper is organized as follows. In Section~\ref{sec:prelim}, we review $\mathcal{Q}_1$ codes and the measurement-based preparation of $\pone$ code-states with error detection from~\cite{goswami2022fault}. In Section~\ref{sec:fact-prep}, we describe our factory preparation, in Section~\ref{sec:estimates_theory}, we provide theoretical estimates of the preparation rate and of the Pauli error probabilities on the prepared state, and in Section~\ref{sec:num-res}, we present our numerical results regarding the factory preparation and comparison with surface codes. Finally, in Section~\ref{sec:conc}, we conclude with some perspectives and future directions.

\section{Preliminaries} \label{sec:prelim}

\subsection{$\pone$ codes}

Here, we briefly review $\pone$ codes, which are CSS quantum polar codes that encode one logical qubit (for a review of CSS quantum polar codes see~\cite[Section~II]{goswami2022fault}).

\smallskip The quantum polar transform $Q_N$, where $N = 2^n$, with $n \geq 0$, is the unitary operation on $N$ qubits that operates in the computational basis as the classical polar transform $P_N$. Precisely, for any $\bm{u} = (u_1, \dots, u_N) \in \{0,1\}^N$, we define $Q_N\ket{\bm{u}} = \ket{P_N\bm{u}}$, where $P_N = \big( \begin{smallmatrix} 1 & 1 \\ 0 & 1 \end{smallmatrix}\big)^{\otimes n} $. Hence, $Q_N$  can be realized by recursively applying the quantum CNOT gate, transversely, on sub-blocks of $2^k$ qubits, for $k = 0,...,n-1$ (see Fig. \ref{fig:qpolar_N8}).

\smallskip Let $ \mathcal{S} = \{1,...,N\}$ denote an $N$-qubit quantum system. For a $\pone$ code, a position $i \in \mathcal{S}$ is chosen to encode the logical information. Given the index $i$, the set of indices preceding $i$,  $i.e$, $\mathcal{Z} \eqdef \{1, \dots, i-1\}$, are frozen in a $Z$ basis state $\ket{\bm{u}}_\mathcal{Z}, \bm{u} \in \{0, 1\}^{1-i}$. Further, the set of indices succeeding $i$,  $i.e$, $\mathcal{X} \eqdef \{i+1, \dots, N\}$, are frozen in a $X$ basis state, $\ket{\bm{\oline{v}}}_\mathcal{X}$, where $\bm{v} \in \{0, 1\}^{|\mathcal{X}|}$ and we have used the notation $\ket{\bar{0}} := \ket{+}$, and $\ket{\bar{1}} := \ket{-}$. 


\smallskip Therefore, the logical code-state, denoted by $\ket{\widetilde{\phi}}_\mathcal{S}$, is given by $\ket{\widetilde{\phi}}_\mathcal{S} = Q_N (\ket{\bm{u}}_\mathcal{Z} \otimes \ket{\phi}_i  \otimes \ket{\bm{\oline{v}}}_\mathcal{X})$. In the following, we shall denote by by $\mathcal{Q}_1(N, i)$ the $\pone$ code of length $N$,  with  information position $ i \in \mathcal{S} = \{1, \dots, N \}$.

It is worth emphasizing that the error correction performance of a $\pone$ code greatly depends on the choice of the information position $i$. For depolarizing channels, the information position providing the best error correction performance, depending on the code-length $N$, was determined by using density evolution in~\cite{goswami2022fault}.

\paragraph*{Shor-$\pone$codes.} When the information position $ i \in \mathcal{S}$ is a power of two, $i.e.$, $i = 2^k, 0 \leq k \leq n$, the corresponding $\pone$ code is a Shor code~\cite[Theorem 1]{goswami2022fault}. The sub-family of Shor codes obtained from $\pone$ codes are referred to as Shor-$\pone$ codes. A Shor-$\pone$ code in general has inferior error correction performance compared to a $\pone$ code of the same code-length and minimum distance as the successive cancellation (SC) decoding of $\pone$ codes is able to decode beyond the minimum distance~\cite{goswami2022fault}.

\begin{figure}[!t]
\,\hfill\input{qcircuits/qone_N8}\hfill\,

\vspace*{0.5em}
\caption{An example of the encoding of $\pone$ codes: the figure shows the encoding of the code $\pone (N = 2^3, i = 5)$, with frozen states $\ket{\bm{u}}_\mathcal{Z} = \ket{0, 0, 0, 0}$ and $\ket{\bm{\oline{v}}}_\mathcal{X} = \ket{+, +, +}$.}
\label{fig:qpolar_N8}
\end{figure}

\subsection{Measurement-based preparation of $Q_1$ code-states}
%
%
%
In this section, we summarize the measurement-based preparation from~\cite{goswami2022fault} and discuss briefly its fault tolerance, under a circuit-level Pauli noise model.
%
%

Consider a $\pone (N, i)$ code, with $N = 2^n$ and $i \in \mathcal{S} = \{1, \dots, N\}$. We consider logical $Z$ and $X$ states of the code $\pone(N, i)$, hence the information position is also frozen in either $Z$ or $X$ basis, accordingly. Therefore, a $\pone(N, i)$ code-state has the following form,
\begin{equation}
\ket{q_N}_\mathcal{S} := Q_N \left( \ket{\bm{u}, \oline{\bm{v}}}_\mathcal{S}\right) = Q_N \left(\ket{\bm{u}}_{\cl{Z}(n)} \otimes \ket{ \oline{\bm{v}}}_{\cl{X}(n)}\right), \label{eq:q-prep-state}
\end{equation}
where $\cl{Z}(n) = \{1, \dots, i(n)\}$ and $\cl{X}(n) = \{i(n) + 1, \dots, N\}$, where \[i(n) = \begin{cases}   
 i, & \text{ for $Z$ logical state.} \\
 i-1, & \text{ for $X$ logical state.}
 \end{cases}\]


\smallskip Therefore, $i(n)$ simply represents the length of $Z$ type frozen set after the $n^{th}$ level of recursion. When no confusion is possible, we may simply write $\ket{q_N}$ instead of $\ket{q_N}_\mathcal{S}$. Finally, $\pone$ states defined by the same value of $i(n)$ are considered equivalent, regardless of the corresponding frozen values $\bm{u}, \bm{v}$. Note that equivalent $\pone$ states are defined by the same stabilizer generators, up to sign factors~\cite[Lemma 4]{goswami2022fault}.

\subsubsection{Recursive measurement-based preparation without noise}
\begin{figure}[!t]
\,\hfill\hspace*{2em}\scalebox{0.95}{\input{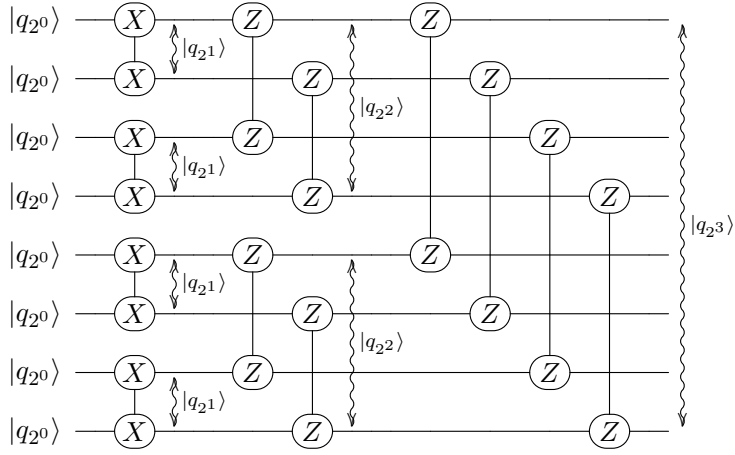}}\hfill\,

\caption{An example of Procedure~\ref{prot:prep}: the figure shows the measurement-based preparation of $\ket{q_N}_\mathcal{S}$ in (\ref{eq:q-prep-state}), with $N = 2^3, i(n) = 3$, that is, $\ket{0}_L$ corresponding to $\pone (N = 2^3, i = 3)$.  Here, slightly flattened circles connected by a~vertical wire denote either an $X \otimes X$ or a $Z \otimes Z$~measurement on the corresponding qubits, and $\ket{q_{2^k}}$ are equivalent $\pone$ states of length $2^k$ ($\ket{q_{2^0}}$ is a Pauli $Z$ basis state).}
\label{fig:qpolarprep_N8_i3}
\end{figure}

Any $\pone$ code-state can be prepared using the following measurement-based procedure~\cite[Theorem~1]{goswami2022fault} (see also Fig.~\ref{fig:qpolarprep_N8_i3}).

\begin{procedure}[measurement-based Preparation \cite{goswami2022fault}] \label{prot:prep}
Consider $\ket{q_N}_\mathcal{S}$ from (\ref{eq:q-prep-state}) and let $b_1\cdots b_n$ be the binary representation of $i(n)-1$, with $b_n$ being the most significant bit, $i.e$, $i(n)-1 = \sum_{k=1}^{n} b_k 2^{k-1}$. Then, the measurement-based procedure to prepare $\ket{q_N}_\mathcal{S}$ is carried out in $n+1$ steps, as follows.

\begin{list}{}{\setlength{\labelwidth}{2em}\setlength{\leftmargin}{1.7em}\setlength{\listparindent}{0em}}

\item[$(0)$] First, $\mathcal{S} = \{1, \dots, N\}$ is initialized in a Pauli $Z$ basis state $\ket{\bm{u}}_{\mathcal{S}}, \text{ for some } \bm{u} \in \{0, 1\}^N$.

\item[$(1 \to n)$] Then, two-qubit Pauli measurements are recursively applied for $n$ levels. The recursion is the same as the recursion of the quantum polar transform (see Fig. \ref{fig:qpolar_N8}), except that each $\cnot$ gate is replaced by either Pauli $X \otimes X$ or $Z \otimes Z$ measurement. Precisely, if $b_k = 0$ (or, $b_k = 1$), we apply Pauli $X \otimes X$ (or, $Z \otimes Z$) measurements at the $k^\text{th}$ recursion level, $k = 1, \dots, n$.
\end{list}
\end{procedure}
%

Note that a Pauli $Z$ basis state can be considered as a $\pone$ code-state of length $2^0$. Therefore, the first step of Procedure \ref{prot:prep}, \emph{i.e.}, initialization in a Pauli $Z$ basis, corresponds to the zeroth level of recursion, where one prepares $2^n$ copies of $\ket{q_{2^0}}$. After any $k^\text{th}$ level of recursion, $1 \leq k \leq n$, Procedure \ref{prot:prep} prepares $2^{n-k}$ equivalent code-states $\ket{q_{2^k}}$, with $i(k)-1 = \sum_{k=1}^{k} b_k 2^{k-1}$ (see Fig.~\ref{fig:qpolarprep_N8_i3}). In particular, each $\ket{q_{2^k}}$ is produced by applying  transversal Pauli $X \otimes X$ or $Z \otimes Z$ measurements on two equivalent $\ket{q_{2^{k-1}}}$ from the $(k-1)^{th}$ level of recursion \cite[Lemma 1]{goswami2022fault}. 


\subsubsection{Recursive measurement-based preparation with noise} \label{sec:prep-n}
We consider the standard implementation of Pauli $Z \otimes Z$ and $X \otimes X$ measurements, using an ancilla qubit, as depicted in Figure~\ref{fig:mZZ_mXX_notation_and_circuit}. Then, the measurement-based procedure consists of the following basic components: qubits initialization in either Pauli $X$ or $Z$ basis, $\cnot$ gates, measurements in the Pauli $X$ or Pauli $Z$ basis. It is easy to see that the total number of components in the preparation of a $\pone$ code-state  of length $N$, denoted here by $C_N$, is given by \cite{goswami2022fault},
\begin{equation} \label{eq:num-comp}
C_N = N (1 + 2 \log N)
\end{equation}

\paragraph*{Noise Model.} We further assume that each component fails independently with some probability $p$, referred to as the physical error rate, according to a circuit level depolarizing noise model as follows~\cite{fowler2012surface}.


\begin{list}{}{\setlength{\labelwidth}{2em}\setlength{\leftmargin}{0.3em}\setlength{\listparindent}{0em}}

\item[$1$] A noisy initialization in Pauli $Z$ (or $X$) basis corresponds to the perfect initialization, followed by an $X$ (or $Z$) error on the initialized qubit, with  probability $p$.

\item[$2$] A noisy CNOT gate corresponds to applying the perfect CNOT gate, followed by a two-qubit depolarizing channel, with probability $p$. Precisely, after the perfect CNOT, any one of the following $15$-two qubit Pauli errors $I \otimes X, I \otimes Y, I \otimes Z, X \otimes I, X \otimes X, X \otimes Y, X \otimes Z$,  $Z \otimes I$, $Z \otimes X$, $Z \otimes Y$, $Z \otimes Z$, $Y \otimes I$, $Y \otimes X$, $Y \otimes Y$, $Y \otimes Z$ may happen, with total probability $p$ (probability $p/15$ each).

\item[$3$] A noisy Pauli $Z$ (or $X$) basis measurement corresponds to first applying a Pauli $X$ (or $Z$) error on the qubit to be measured with probability $p$, then doing the perfect Pauli $Z$ (or $X$) measurement.

\end{list}

\begin{figure}[!t] 
\centering
\begin{subfigure}[t]{.9\linewidth}
\begin{subfigure}[t]{.4\linewidth}
\,\hfill\input{qcircuits/measureZZ_notation}\,
\end{subfigure}\hfill%
\raisebox{-1.4em}{$\equiv$}\hfill%
\begin{subfigure}[t]{.5\linewidth}
\,\qquad\input{qcircuits/measureZZ_circuit}\hfill\,
\end{subfigure}
\captionsetup{justification=centering}
\caption{Pauli $Z \otimes Z$ measurement}
\label{fig:mZZ_notation_and_circuit}
\vspace*{3mm}
\end{subfigure}
\begin{subfigure}[t]{.9\linewidth}
\begin{subfigure}[t]{.4\linewidth}
\,\hfill\input{qcircuits/measureXX_notation}\,
\end{subfigure}\hfill%
\raisebox{-1.4em}{$\equiv$}\hfill%
\begin{subfigure}[t]{.5\linewidth}
\,\qquad\input{qcircuits/measureXX_circuit}\hfill\,
\end{subfigure}
\captionsetup{justification=centering}
\caption{Pauli $X \otimes X$ measurement}
\label{fig:mXX_notation_and_circuit}
\end{subfigure}
\caption{Two-qubit Pauli measurements:    Figures (a) and (b) provide quantum circuits implementing Pauli $Z \otimes Z$ and $X \otimes X$ measurements, respectively. }
\label{fig:mZZ_mXX_notation_and_circuit}
\end{figure}



\smallskip The measurement-based preparation given in Procedure~\ref{prot:prep} is not fault-tolerant by itself under the above noise model. The reason is that the outcomes of transversal Pauli $Z \otimes Z$ or Pauli $X \otimes X$ measurements, which are needed to determine the values corresponding to the frozen $\mathcal{Z}$ and $\mathcal{X}$ sets, get error corrupted, thus leading to a wrong determination of frozen values~\cite{goswami2022fault}. 

\smallskip To achieve fault tolerance, an error detection gadget is incorporated into the measurement-based procedure. Taking advantage of the redundancy in the transversal Pauli measurements, the error detection gadget detects errors at each level of recursion.

\paragraph{Error detection gadget.} Here, we briefly present the error detection gadget for the case of Pauli $Z \otimes Z$ measurements  (see~\cite[Procedure 2]{goswami2022fault} for more details). For $K = 2^k$, consider two equivalent $\pone$ code states $\ket{q^1_{K/2}}_{\mathcal{S}_1} = Q_{K/2}\ket{\bm{u}_1, \oline{\bm{v}_1}}$ and $\ket{q^1_{K/2}}_{\mathcal{S}_2} = Q_{K/2}\ket{\bm{u}_2, \oline{\bm{v}_2}}$, where $\bm{u}_1, \bm{u}_2 \in \{0, 1\}^{i(k-1)}$ and $\bm{v}_1, \bm{v}_2 \in \{0, 1\}^{K - i(k-1)}$. Consider first the noiseless scenario. The result of the transversal Pauli- $Z \otimes Z$ measurements on these polar code states is a codeword of a classical polar code as follows,
\begin{equation}
 \bm{m} = P_\frac{K}{2}(\bm{u'}, \bm{x}) \in \{0, 1\}^{\frac{K}{2}}, \label{eq:m-out}
\end{equation}
where $\bm{u'} = \bm{u_1} \oplus \bm{u_2} \in \{0, 1\}^{i(k-1)}$ and  $\bm{x} \in \{0, 1\}^{\frac{K}{2}-i(k-1)}$ is a random unknown vector, and $P_{\frac{K}{2}}$ is the classical polar transform. After measurements, the state of the joint system $\mathcal{S} = \mathcal{S}_1 \cup \mathcal{S}_2$ is a $\pone$ state, 
\begin{equation}
\ket{q_K}_\mathcal{S} =  Q_K \ket{(\bm{u'}, \bm{x}, \bm{u_2}), \overline{\bm{v_1} \oplus \bm{v_2}}}_\mathcal{S},
\end{equation}
 with $i(k) = i(k-1) + K/2 > K/2$, and where $\bm{x}$ is determined from the measurement outcome $\bm{m}$ in (\ref{eq:m-out}) by, $\bm{x} = P_{\frac{K}{2}}(\bm{m})\lvert_{\mathcal{X}(k-1)}$, $i.e.$, the subvector of  $P_{\frac{K}{2}}(\bm{m}) \in \{0, 1\}^{K/2}$ corresponding to indices in the set $\mathcal{X}(k-1)$.

For the noisy scenario, the measurement outcome gives a noisy codeword of the classical polar code instead of (\ref{eq:m-out}), as follows
\begin{equation}
\bm{m} = P_\frac{K}{2}(\bm{u'}, \bm{x}) \oplus \bm{e}_X.
\end{equation}
The error detection gadget determines the syndrome of the error term $\bm{e}_X$ in the measurement outcome $\bm{m}$ as, $P_{\frac{K}{2}}(\bm{e}_X) \lvert_{\mathcal{Z}(k-1)} = P_{\frac{K}{2}}(\bm{m})\lvert_{\mathcal{Z}(k-1)} \,\oplus\, \bm{u'}$. If the syndrome is zero, we proceed as in the noiseless case. If the syndrome is not zero, we report a component failure and discard the prepared state. For example, in Fig.~\ref{fig:qpolarprep_N8_i3}, if an error is detected in one of the $\ket{q_2^2}$ prepared at the second level of recursion, we discard the other prepared state even if no error is detected in it and restart the procedure from the beginning.


\medskip Therefore, one needs to repeat the preparation until a preparation succeeds without an error detection. The preparation rate is defined as 
\begin{equation} \label{eq:prep_rate}
p_{\text{prep}}= \lim_{R \to \infty} \frac{t}{R},
\end{equation}
where $t$ is the number of successful preparations out of $R$ independent preparation attempts. 

\smallskip Finally, it is worth noting that some errors may not be detected by the gadget and hence, will remain on the successfully prepared states. It has been shown that the successfully prepared state is fault-tolerant in the sense that the errors in the prepared state do not exceed the number of component failures~\cite[Theorem 3]{goswami2022fault}.

\section{A factory preparation of polar code-states} \label{sec:fact-prep}


 

Note that the rate of the preparation $p_{\text{prep}}$ in~(\ref{eq:prep_rate}) decreases as the code-length $N$ increases. Intuitively, this is because the number of components increases with respect to $N$, as given in~(\ref{eq:num-comp}), thus resulting in an increased expected number of failures, and as a consequence a higher probability of error detection. Numerical results in~\cite{goswami2022fault} suggest that the $p_{\text{prep}}$ decreases rapidly as $N$ increases, hence, prohibiting the preparation of $\pone$ code-states of larger code-lengths in a practical scenario.


In this section, to improve the preparation rate, we consider a factory preparation, where several $\pone$ code-states of length $N$ are prepared in parallel.  We modify~\cite[Procedure 2]{goswami2022fault}, so that at some intermediate levels of recursion, we only discard $\pone$ code-states where an error is detected, and keep all the successfully prepared intermediate states and continue the preparation for the next levels of recursion. As this allows to avoid restarting the preparation from the beginning every time an error is detected, we may achieve better preparation rate than~\cite[Procedure 2]{goswami2022fault}.



\medskip We below describe our factory preparation in detail.

Let $\mathcal{S}_T \eqdef \{1, \dots, TN \}, T \geq 1$ be a set of $TN$ qubits, on which we want to prepare several copies of a $\pone$ code-state of length $N = 2^n$, $i.e.$, $\ket{q_N}$, with given $i(n)$ value, so that $b_1\cdots b_n \in \{0, 1\}^n$ is the binary expansion of $i(n)-1$. We will refer to $T$ as the \emph{size of the factory}, as $T$ is the maximum number of copies of $\ket{q_N}$ that can be produced. Further, consider the following ordered set
\begin{equation} \label{eq:sch-set}
    n_{sch} := \{ i_1, i_2, ..., i_{|n_{sch}|}\} \subseteq \{1, ..., n \},
\end{equation}
such that $0 < i_1 < i_2 < \cdots < i_{|n_{sch}|}$, and $i_{|n_{sch}|} = n$. We shall refer to $n_{\text{sch}}$ as the \emph{scheduling set} and the elements in $n_{\text{sch}}$ as the \emph{scheduling recursion levels}.  

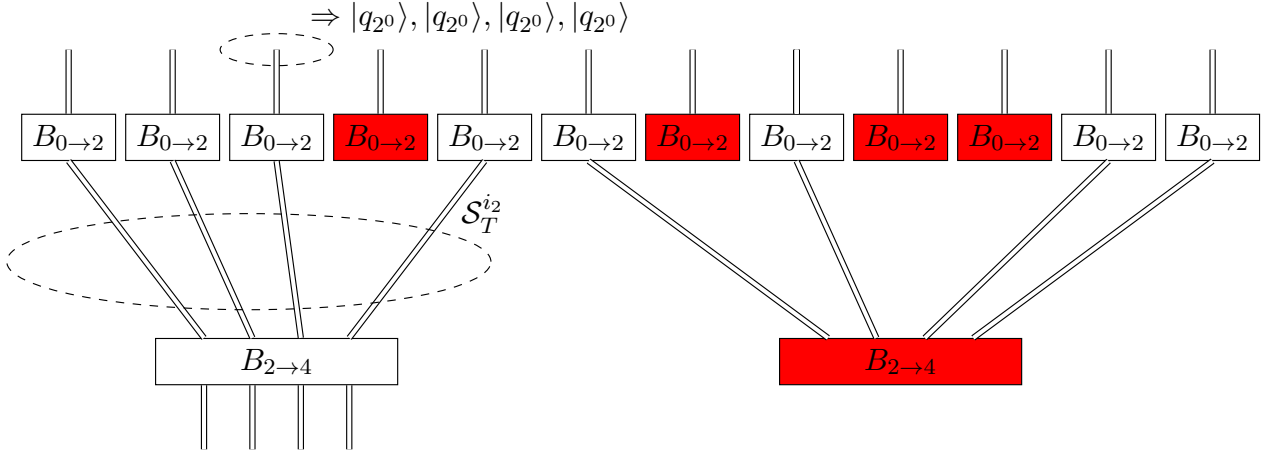
\begin{figure}[!t]
    \centering
    \resizebox{\linewidth}{!}{%
    \begin{tikzpicture}
        \draw 
            (0, 0) node[draw, minimum width=1.0cm, minimum height = 0.5cm] (B1) {$B_{0 \to 2}$}
            (B1.east) ++(0.7, 0) node[draw, minimum width=1.0cm, minimum height = 0.5cm] (B2) {$B_{0 \to 2}$}
            (B2.east) ++(0.7, 0) node[draw, minimum width=1.0cm, minimum height = 0.5cm] (B3) {$B_{0 \to 2}$}
            (B3.east) ++(0.7, 0) node[draw, minimum width=1.0cm, minimum height = 0.5cm, fill = red] (B4) {$B_{0 \to 2}$}
            (B4.east) ++(0.7, 0) node[draw, minimum width=1.0cm, minimum height = 0.5cm] (B5) {$B_{0 \to 2}$}
            (B5.east) ++(0.7, 0) node[draw, minimum width=1.0cm, minimum height = 0.5cm] (B6) {$B_{0 \to 2}$}
            (B6.east) ++(0.7, 0) node[draw, minimum width=1.0cm, minimum height = 0.5cm, fill = red] (B7) {$B_{0 \to 2}$}
            (B7.east) ++(0.7, 0) node[draw, minimum width=1.0cm, minimum height = 0.5cm] (B8) {$B_{0 \to 2}$}
            (B8.east) ++(0.7, 0) node[draw, minimum width=1.0cm, minimum height = 0.5cm, fill = red] (B9) {$B_{0 \to 2}$}
            (B9.east) ++(0.7, 0) node[draw, minimum width=1.0cm, minimum height = 0.5cm, fill = red] (B10) {$B_{0 \to 2}$}
            (B10.east) ++(0.7, 0) node[draw, minimum width=1.0cm, minimum height = 0.5cm] (B11) {$B_{0 \to 2}$}
            (B11.east) ++(0.7, 0) node[draw, minimum width=1.0cm, minimum height = 0.5cm] (B12) {$B_{0 \to 2}$}
            ;
        \draw
            ($0.5*(B1) + 0.5*(B5)$) ++ (0, -2.8) node[draw, minimum width=3.0cm, minimum height = 0.5cm] (C1) {$B_{2 \to 4}$}
            ($0.5*(B6) + 0.5*(B12)$) ++ (0, -2.8) node[draw, minimum width=3.0cm, minimum height = 0.5cm, fill = red] (C2) {$B_{2 \to 4}$}
            ;
        \draw[double, double distance = 0.5mm] 
            (B1.north) to ++(0, +0.8)
            (B2.north) to ++(0, +0.8)
            (B3.north) to ++(0, +0.8)
            (B4.north) to ++(0, +0.8)
            (B5.north) to ++(0, +0.8)
            (B6.north) to ++(0, +0.8)
            (B7.north) to ++(0, +0.8)
            (B8.north) to ++(0, +0.8)
            (B8.north) to ++(0, +0.8)
            (B9.north) to ++(0, +0.8)
            (B10.north) to ++(0, +0.8)
            (B11.north) to ++(0, +0.8)
            (B12.north) to ++(0, +0.8)
            (C1.north) ++(-0.9, 0) to (B1.south)
            (C1.north) ++(-0.3, 0) to (B2.south)
            (C1.north) ++(+0.3, 0) to (B3.south)
            (C1.north) ++(+0.9, 0) to (B5.south)
            (C2.north) ++(-0.9, 0) to (B6.south)
            (C2.north) ++(-0.3, 0) to (B8.south)
            (C2.north) ++(+0.3, 0) to (B11.south)
            (C2.north) ++(+0.9, 0) to (B12.south)
            (C1.south) ++(-0.9, 0) to ++(0, -0.8)
            (C1.south) ++(-0.3, 0) to ++(0, -0.8)
            (C1.south) ++(+0.3, 0) to ++(0, -0.8)
            (C1.south) ++(+0.9, 0) to ++(0, -0.8)
            ;
        \draw
            ($0.5*(C1.north) + 0.5*(B2)$) node[](f1){}
            (f1) ++ (-1.5, 0)node[](f2){}
            (f1) ++ (+2.5, 0)node[](f3){}
            ;
        \draw[dashed]
            (f1)++(0.3, -0.3) ellipse (3cm and 0.6cm) ++(2.9, 0.6)node[](){$\mathcal{S}^{i_2}_T$}
            (B3.north) ++ (0, 0.8) ellipse (0.7cm and 0.2cm) ++(2.4, 0.4) node[](){$ \Rightarrow \ket{q_{2^0}}, \ket{q_{2^0}}, \ket{q_{2^0}}, \ket{q_{2^0}}$}
            ;
        
    \end{tikzpicture}
    }
    \caption{An example of Procedure~\ref{prot:fact_prep}: the figure illustrates the factory preparation for $N = 16, T = 3$ and $n_{\text{sch}} = \{0, 2, 4\}$, successfully preparing one copy of $\ket{q_N}$.  The factory preparation proceeds from top to bottom. Each double line denotes a set of four qubits and the box $B_{i \to j}$, $0 \leq i < j \leq n$ denotes the preparation using~\cite[Procedure 2]{goswami2022fault} between $(i+1)^{th}$ and $j^{th}$ level of recursion, and a red box $B_{i \to j}$ signifies that an error was detected, and therefore, no output is produced. Firstly, we initialize qubits in $\mathcal{S}_T$ (\emph{i.e.}, the set of all qubits) in the Pauli $Z$ basis. Then, we split the set $\mathcal{S}_T$ in groups containing four qubits, which is denoted by the double wires in the figure. Then, on each group, we attempt to prepare the code-state $\ket{q_{2^2}}$, by applying $B_{0 \to 2}$ (note that for the analysis in Section~\ref{sec:estimates_theory}, we actually consider the initialization step as part of $B_{0 \to 2}$). Further, we split the successful prepared states into groups, each containing four copies of $\ket{q_{2^2}}$. On each group, we attempt to prepare $\ket{q_{2^4}}$, by applying $B_{2 \to 4}$.} 
    \label{fig:fact_prep}
\end{figure}

\smallskip We denote by $B_{i \to j}$ the recursion levels from $i+1,...,j$ of Procedure 1, with respect to the binary string $b_{i+1},...,b_{j}$. For the particular case $B_{0 \to j}$, we  also include the initialization of the data qubits. We shall also assume that $B_{i \to j}$ incorporates the error detection gadget, as explained in Section~\ref{sec:prep-n}, and detailed in~\cite[Procedure 2]{goswami2022fault}. If an error is detected at one of the recursion levels of $B_{i \to j}$, we declare a preparation failure and discard the prepared state, hence no output is produced. If no error is detected during $B_{i \to j}$, we successfully prepare a $\pone$ code state of length $2^{j}$ (as shown in Fig.~\ref{fig:fact_prep}).

%

\smallskip
Our factory preparation is as follows (see also Fig.~\ref{fig:fact_prep}). 

\begin{procedure}[Factory Preparation] \label{prot:fact_prep}

Consider the set of qubits $\mathcal{S}_T = \{1, \dots, TN\}, T \geq 1, N = 2^n, n > 0 $, and the scheduling set $n_{\text{sch}} \subseteq \{1, \dots, n\}$ according to~(\ref{eq:sch-set}). Then, the factory preparation consists of the following steps.

\begin{list}{}{\setlength{\labelwidth}{2em}\setlength{\leftmargin}{2.5em}\setlength{\listparindent}{0em}}

\item[$(i)$] We first split $\mathcal{S}_T$ in groups, each containing $2^{i_1}$ qubits. We then apply $B_{0 \to i_1}$ on each group.


\item[$(ii)$] To prepare several copies of $\ket{q_{2^n}}$, we then recursively apply the preparation $B_{i_k \to i_{k+1}}, 1 \leq k < |n_{sch}|$,  as follows.

After any scheduling recursion level $i_k$, $1 \leq k < |n_{sch}|$, if more than $2^{n - i_k}$ copies of $\ket{q_{2^{i_k}}}$ are successfully prepared, we do the following. 

We split the set of successfully prepared code-states $\ket{q_{2^{i_k}}}$ into groups, each containing $2^{i_{k+1} - i_k}$ copies of $\ket{q_{2^{i_k}}}$. We then attempt to prepare $\ket{q_{2^{i_{k+1}}}}$ by applying $B_{i_k \to i_{k+1}}$ on each group. 

\vspace*{0.6em}
Otherwise, if less than $2^{n - i_k}$ copies of $\ket{q_{2^{i_k}}}$ are successfully prepared, we declare a preparation failure and discard entirely the factory preparation.

\end{list}
\end{procedure}

\smallskip Note that after a scheduling recursion level $i_k$, we need to have at least $2^{n - i_k}$ copies of $\ket{q_{2^{i_k}}}$ to be able to prepare at least one copy of $\ket{q_{2^{n}}}$. Therefore, if the number of successfully prepared state is less than $2^{n - i_k}$, we discard the factory preparation. In other words, the factory preparation is successful if we have at least $2^{n - i_k}$ successfully prepared copies of $\ket{q_{2^{i_k}}}$, after all the scheduling recursion levels $i_k, 1 \leq k \leq |n_{sch}|$.

\smallskip Finally, note that for $n_{sch} = \{0, n\}$, the factory preparation corresponds to applying $B_{0 \to n}$, hence it is the same as the preparation~\cite[Procedure 1]{goswami2022fault} with error detection.

\subsection{Preparation rate and error probabilities of factory preparation}
Consider the factory preparation of $\ket{q_N}$ from Procedure~\ref{prot:fact_prep}, with respect to some $T \geq 1$ and $n_{\text{sch}} \subseteq \{1, \dots, n \}$.  Suppose we run the factory preparation $R$ times, successfully preparing $t_R \geq 0$ copies of $\ket{q_N}$ in total. Then, we define the preparation rate of the factory preparation, denoted by $p^{T, n_{\text{sch}}}_{\text{fact}}$, as follows,
\begin{equation} \label{eq:fact_prep}
p^{T, n_{\text{sch}}}_{\text{fact}} := \lim_{R \to \infty} \frac{t_R}{RT}.
\end{equation}


Note that for $T = 1$, $p^{T, n_{\text{sch}}}_{\text{fact}}$ is equal to $p_{\text{prep}}$ from (\ref{eq:prep_rate}), $i.e.$, the preparation rate of~\cite[Procedure 2]{goswami2022fault}. Another interesting case is when $T \to \infty$, for which we define,
\begin{equation} \label{eq:fact_prep_T}
p^{n_{\text{sch}}}_{\text{fact}} \eqdef \lim_{T \to \infty} p^{T, n_{\text{sch}}}_{\text{fact}}.
\end{equation}

\smallskip Let $\bm{e}^j_X, \bm{e}^j_Z \in \{0, 1\}^N$ be the $X$ and $Z$ errors, respectively, on the $j^{th} \in \{1, \dots, t_R\}$ successfully prepared state $\ket{q_N}$. Then, the average $X$ and $Z$ error probabilities, denoted respectively by $p^{prep}_X$ and $p^{prep}_Z$, are as follows
%
%
\begin{align}
p^{prep}_X &=  \lim_{R \to \infty} \frac{1}{t_RN} \sum_{j=1}^{t_R} \wt(\bm{e}^j_X), \label{eq:avg_X} \\ 
p^{prep}_Z &= \lim_{R \to \infty} \frac{1}{t_RN} \sum_{j = 1}^{t_R} \wt(\bm{e}^j_Z), \label{eq:avg_Z} 
\end{align}
where $\wt(\bm{e}^j_X)$ is the Hamming weight of $\bm{e}^j_X \in \{0, 1\}^N$. From now on, we refer to $p^{prep}_X$ and $p^{prep}_Z$ as the $X$ and $Z$ \emph{preparation error probabilities}, respectively. The preparation error probabilities may be used to estimate the logical error rate of $\pone$ code under  Steane's error correction, with the help of the density evolution technique as in~\cite{goswami2022fault}.



\section{Theoretical estimates of preparation rate and error probabilities} \label{sec:estimates_theory}



In this section, we define the notions of rough and smooth\footnote{ The terms ``smooth" and ``rough" are used to distinguish $X$ and $Z$ boundaries in the topological quantum code literature. We emphasize that our definition is unrelated to the one used for topological codes.} errors for the measurement based preparation, and then using them, we provide theoretical estimates of the preparation rate $p_{prep}$ from~(\ref{eq:fact_prep_T}) and preparation error probabilities $p^{prep}_X$ and $p^{prep}_Z$ from~(\ref{eq:avg_X}) and~(\ref{eq:avg_Z}), respectively.

\subsection{Rough and smooth errors with respect to $B_{i \to j}$}

\smallskip Recall that $B_{i \to j} $ corresponds to the recursion levels $i+1, \dots, j$ of the measurement based procedure with respect to the binary string $b_{i+1}, \dots, b_{j}$, where $b_k = 0$ signifies a recursion level with Pauli $X \otimes X$ measurements and $b_k = 1$  a recursion level with Pauli $Z \otimes Z$ measurements. For $i = 0$, $B_{i \to j}$ also includes initialization in Pauli $Z$ basis at the zeroth level of recursion. Further, $B_{i \to j}$ takes as input $2^{j-i}$ copies of $\ket{q_{2^i}}$, and produces as output a copy of $\ket{q_{2^j}}$ if no error is detected. 

\smallskip Consider the set of data qubits $\mathcal{S}_{i \to j} \eqdef \{1, \dots, 2^{j}\}$ that are input to $B_{i \to j}$, and let $\mathcal{A}_{ i \to j} := \{1, \dots, 2^{j-1} \}$ be the set of ancilla qubits used to implement the Pauli $Z \otimes Z$ and Pauli $X \otimes X$ measurements in $B_{i \to j}$. Further, let $\mathcal{C}_{i \to j}$ be the set of all components corresponding to $B_{i \to j}$. Note that $\mathcal{C}_{i \to j}$ consists of $T_{i \to j} = \big(j - i\big) 2^{j-1}$ two qubit Pauli measurements. Therefore, it consists of $T_{i \to j}$ initializations and measurements of the ancilla qubits and $2T_{i \to j}$ $\cnot$ gates between the  data and ancilla qubits. If $i = 0$, it also has $2^j$ initializations in the Pauli $Z$ basis.

\begin{figure}[!t] 
\centering
\,\hfill\hspace*{2em}\input{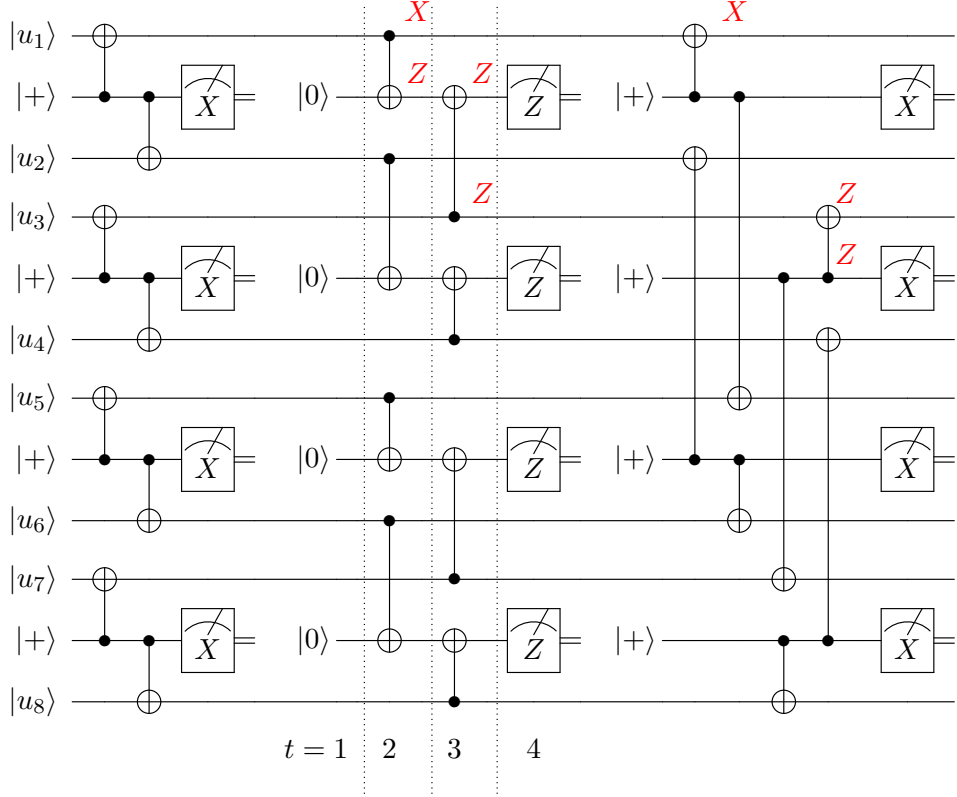}\hfill\,
\caption{An example of rough error: the figure shows a faulty measurement-based preparation (Procedure~\ref{prot:prep}) of $\ket{0_L}$ corresponding to the code $\pone(N = 8, i(n) = 2)$, where Pauli $Z \otimes Z$ and Pauli $X \otimes X$ measurements are implemented according to the circuits in Fig.~\ref{fig:mZZ_mXX_notation_and_circuit}. A $\cnot$ failure at $k = 2, t= 2$ produces the non-trivial error $X \otimes Z$, which flips the measurement outcome of a Pauli $X$ measurement at $k = 3, t= 4$. Therefore, the produced error is a rough error.}
\label{fig:error_det_big}
\end{figure}

\medskip Any recursion level of $B_{i \to j}$ consists of the following four~\emph{time steps} (see also Fig.~\ref{fig:error_det_big}),

\begin{list}{}{\setlength{\labelwidth}{2em}\setlength{\leftmargin}{2.5em}\setlength{\listparindent}{0em}}

\item[$(t = 1)$] The ancilla qubits in $\mathcal{A}_{ i \to j}$ are initialized in the Pauli $Z$ or $X$ basis.

\item[$(t = 2)$] The first $\cnot$ gate (corresponding to all Pauli  $X \otimes X$ or $Z \otimes Z$ measurements) are applied in parallel.

\item[$(t = 3)$] The second $\cnot$ gate (corresponding to all Pauli $X \otimes X$ or $Z \otimes Z$ measurements) are applied in parallel.

\item[$(t = 4)$] The ancilla qubits in $\mathcal{A}_{ i \to j}$ is measured in the Pauli $Z$ or $X$ basis.
 
\end{list}

%
%
%
%

%

\smallskip We now define rough and smooth errors with respect to $B_{i \to j}$, as follows.
\begin{definition}[Rough and smooth errors] \label{def:eff_err}
Let $\ket{\psi_{i \to j}^{k,t}}$ be the quantum state corresponding to the joint system $\mathcal{S}_{i \to j} \cup \mathcal{A}_{i \to j}$, after a time step $t = 1, 2, 3, 4$ of the $k^{th}$ recursion level of $B_{i \to j}$ ($ i < k \leq j  $). We say a Pauli error $P_e$ acting on $\ket{\psi_{i \to j}^{k,t}}$ is a rough error if it satisfies the following two conditions:
%
%
%
%
\begin{list}{}{\setlength{\labelwidth}{2em}\setlength{\leftmargin}{2.5em}\setlength{\listparindent}{0em}}

\item[$(C.1)$] It is a non-trivial error in the sense that it not a stabilizer of the quantum state $\ket{\psi_{i \to j}^{k,t}}$,

\item[$(C.2)$] It flips the outcome of at least one single qubit Pauli $Z$ or $X$ measurement at a recursion level $ k', k \leq k' \leq j $.

\end{list}
Further, we say that $P_e$ is a smooth error if it satisfies the above condition $(C.1)$, however it does not satisfy the condition $(C.2)$.
\end{definition}

\smallskip  The condition $(C.2)$ of Definition~\ref{def:eff_err} is illustrated in Fig.~\ref{fig:error_det_big} for $B_{0 \to 3}$, where the error $P_e = X \otimes Z$ happens at $k = 2, t= 2$. Recall that an $X$ error propagates through the control of a $\cnot$ to its target, while it simply passes through the target. An $Z$ error propagates through the target of a $\cnot$ gate to its control, while it simply passes through the control. Further, an error flips the outcome of a Pauli measurement if they anti-commute with each other. The error $P_e$ in Fig.~\ref{fig:error_det_big}, consisting of an $X$ error on the first data qubit and a $Z$ error on the first ancilla qubit, flips the outcome of an $X$ measurement at $k=3, t = 4$. Since $p_e$ is not a stabilizer of $\ket{\psi_{0 \to 3}^{2,2}}$, it corresponds to a rough error.

The rough and smooth errors are related with the error detection gadget as follows. An error is detected only if it flips the outcome of at least one measurement. Hence, roughness is a necessary condition for error detection. However, it is not a sufficient condition as an error of large weight may flip several measurement outcomes and the error detection is limited by the minimum distance of the classical code. Nevertheless, due to recursive nature of the preparation procedure, for an error to survive it should not be detected at any of the recursion levels that follow. Therefore, we expect that the rough errors will go undetected with small probabilities.  This justifies the following assumption that we make for the estimation of the preparation rate and preparation error probabilities. 

\begin{assumption} \label{ass:err_det}
Any rough error on the quantum state $\ket{\psi^{k,t}_{i \to j}}, i < k \leq j, t = 1, 2, 3,4$, is detected by the error detection gadget at one of the recursion levels $k, \dots, j$.
\end{assumption}




%
%
%
We will further use the rough error probability and smooth error channel, defined below.

\begin{definition}[Rough error probability] 
We define the rough error probability for a component $C \in \mathcal{C}_{i \to j}$, as the probability that $C$ produces a rough error in $B_{i \to j}$.
\end{definition} 
Considering Assumption~\ref{ass:err_det}, the error that remains on the output of $B_{i \to j}$, when no error is detected, is due to the smooth errors. We  define below the smooth error channel corresponding to a component. 
\begin{definition}[Smooth error channel] 
Let $L(H_{\mathcal{S}_{ i \to j}})$ be the set of linear operators acting on $H_{\mathcal{S}_{ i \to j}}$, the Hilbert space corresponding to $\mathcal{S}_{ i \to j}$. Then, we define the smooth error channel corresponding to a component $C \in \mathcal{C}_{i \to j}$, denoted by $\mathcal{W}_C: L(H_{\mathcal{S}_{ i \to j}}) \to L(H_{\mathcal{S}_{ i \to j}})$, as the channel that acts on the output quantum state of $B_{i \to j}$, due to the smooth errors produced by the faults in $C$. 
\end{definition} 
In the following, we denote by $p_C$ and $\mathcal{W}_C$, the rough error probability and the smooth error channel associated with a component $C$, respectively. We further define $k^{\min}_{i \to j}$, as follows,
%
%
\begin{equation}
k^{\min}_{i \to j} := \min \big\{ k \in \{i+1, \dots, j-1\} \mid b_{i'} = b_j, \forall k < i' \leq j \big\}. 
\end{equation}
In other words, $k^{\min}_{i \to j}$ is the minimum value in $\{i+1, \dots, j-1\}$, so that the recursion levels of $B_{i \to j}$ after $k^{\min}_{i \to j}$ consists of only one type of two qubit Pauli measurements, given by the value of $b_j$. 

%

\smallskip We provide below $p_C$ and $\mathcal{W}_C$ for the initialization of data qubits (Lemma~\ref{lem:int-eff-err}), initialization and measurement of ancila qubits (Lemma~\ref{lem:p-w-int-m}), and CNOT gates (Lemma~\ref{lem:p-w-cnot}), considering the circuit level depolarizing noise model from Section~\ref{sec:prep-n}, with the physical error rate $p$. The proofs of lemmas are given in Appendix~\ref{app:proof-lemmas}. 
\begin{lemma} \label{lem:int-eff-err}
Consider the recursion levels corresponding to $B_{0 \to j}, j > 0$ and let $C$ be an initialization component on a data qubit $q \in \mathcal{S}_{0 \to j}$, at the zeroth level of recursion. Then, $p_C$ and $\mathcal{W}_C$ are as follows,
\begin{equation} \label{eq:eff_int_zero}
p_C = \begin{cases}
0, & \text{ if $\sum_{t=1}^j b_t = 0$. } \\
p, & \text{ otherwise. }
\end{cases}
\end{equation}
\begin{equation} \label{eq:non_eff_int_zero}
\mathcal{W}_C = \begin{cases}
I_{\mathcal{S}_{0 \to j} \setminus {q} } \otimes  \mathcal{B}_{q}^{p},  & \text{ if $\sum_{t=1}^j b_t = 0$.} \\
I_{\mathcal{S}_{0 \to j}}, & \text{ otherwise, }
\end{cases}
\end{equation}
where $\mathcal{B}_{q}^{(p)}$ is a bit-flip channel, acting on $q$ with the error probability $p$, i.e., $\mathcal{B}_{q}^{(p)}(\rho) := (1 - p) \rho + p X \rho X$.

\end{lemma}

\begin{lemma} \label{lem:p-w-int-m}
For an initialization and measurement component on an ancilla qubit in $\mathcal{A}_{i \to j}, 0 \leq i < j$, we have the following,
\begin{align}
p_C &= p. \label{eq:eff_measurement}\\
\mathcal{W}_C &= I_{\mathcal{S}_{i \to j}}.
\end{align} 
\end{lemma}
\begin{lemma} \label{lem:p-w-cnot}
Consider a $\cnot_{q \to a}$ on the $k^{th}$ recursion level of $B_{i \to j}, 0 \leq i  < k \leq j$, acting between a data qubit $q \in \mathcal{S}_{i \to j}$ and an ancilla qubit $a \in \mathcal{A}_{i \to j} $. Then,  depending on $k$, $p_C$  is as follows. 
\begin{equation} \label{eq:eff_err_cnot}
 p_C = \begin{cases}
 8p/15, & \text{ if $k =  j$. }  \\
 4p/5, & \text{ if $  k^{\min}_{i \to j} \leq k <  j$.} \\
 14p/15, & \text{ otherwise.} \\
\end{cases}
\end{equation}
Further, $\mathcal{W}_C = I_{\mathcal{S}_{i \to j} \setminus {q}} \otimes \mathcal{W}_q$, where $\mathcal{W}_q$ is a quantum channel acting on $q$ as follows, 
\begin{equation} \label{eq:eff_noise_cnot}
\mathcal{W}_q = \begin{cases}
\mathcal{D}_{q}^{(6p/15)}, & \text{ if $k =  j$ }  \\
 \mathcal{B}_{q}^{(2p/15)}, & \text{ if $k^{\min}_{i \to j} \leq k <  j$, and $b_j = 0$.} \\
  \mathcal{P}_{q}^{(2p/15)}, & \text{if $k^{\min}_{i \to j} \leq k < j$, and $b_j = 1$.} \\
 I_{\mathcal{S}_{i \to j}}, & \text{ otherwise,} \\
\end{cases}
\end{equation}
where $\mathcal{D}_{q}^{(p)}$ is a depolarizing channel, acting on $q$ with error probability $p$ as 
$\mathcal{D}_{q}^{(p)}(\rho) := (1 - p) \rho + \frac{p}{3} (X \rho X + Y \rho Y + Z \rho Z)$, $\mathcal{B}_{q}^{(p)}$ is a bit-flip channel, acting on $q$ with the error probability $p$ as $\mathcal{B}_{q}^{(p)}(\rho) := (1 - p) \rho + p X \rho X$, and $\mathcal{P}_{q}^{(p)}$ is a phase-flip channel, acting on $q$ with the error probability $p$ as $\mathcal{P}_{q}^{(p)}(\rho) := (1 - p) \rho + p Z \rho Z$.
\end{lemma}

\subsection{The success probability of $B_{i \to j}$} \label{sec:succ_prob_Bij}

Considering  Assumption~\ref{ass:err_det}, the success probability of $B_{i \to j}$ corresponds to the probability that a rough error is not produced at any of the recursion levels in $B_{i \to j}$. In this regard, we consider the following two cases: 
\begin{list}{}{\setlength{\labelwidth}{1em}\setlength{\leftmargin}{2.5em}\setlength{\listparindent}{0em}}
\item[$(1)$] The component failures at the previous recursion levels, $\emph{i.e.}$, $B_{0 \to i}$, producing smooth errors with respect to $B_{0 \to i}$.

\item[$(2)$] The component failures at one of the recursion levels corresponding to $B_{i \to j}$.

\end{list}
%
%
Let $p^{(1)}$ and $p^{(2)}$ be the probability that a rough error with respect to $B_{i \to j}$ is not produced due to Point $(1)$ and Point $(2)$, respectively. In the following, we give approximations of $p^{(1)}$ and $p^{(2)}$, using which we may approximate the success probability of $B_{i \to j}$.



\medskip We first consider Point $(2)$. We may approximate $p^{(2)}$ directly from Lemmas~\ref{lem:int-eff-err},~\ref{lem:p-w-int-m} and~\ref{lem:p-w-cnot}, as follows,
%
\begin{equation} \label{eq:p_success_ii}
p^{(2)}  \approx \prod_{C \in \mathcal{C}_{i \to j}}(1-p_C).
\end{equation}
Recall that $p_C$ is the rough error probability associated with the component $C$. We have the following remark.

\begin{figure}[!t]
\,\hfill\input{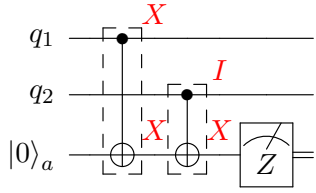} \hfill\,
\caption{ Two rough errors together may produce a smooth error: the figure shows a faulty Pauli $Z \otimes Z$ measurement. The fault in $\cnot_{q_1 \to a}$ produces the error  $X_{q_1} \otimes X_{a}$ and the fault in $\cnot_{q_2 \to a}$ produces the error  $I_{q_2} \otimes X_{a}$. Both errors  $X_{q_1} \otimes X_{a}$ and $I_{q_2} \otimes X_{a}$ are rough  individually as they will flip the measurement outcome of the Pauli $Z$ measurement on the ancilla qubit. However, when these errors are considered together, they do not flip the measurement outcome as the $X$ errors on the ancilla qubit get cancelled. Hence, the total error may not be a rough error. Note that after the measurement, the remaining error on data qubits is $X_{q_1} \otimes I_{q_2}$, which may flip a measurement outcome at one of the subsequent levels of recursion.}
\label{fig:rough-smooth}
\end{figure}

\begin{remark} \label{rem:rough-smooth-err}
We note that the total error produced by component failures in a set $\mathcal{C}' \subseteq \mathcal{C}_{i \to j}$ (containing two or more components, \emph{i.e.}, $|\mathcal{C}'| \geq 2$) can be a smooth error, even though individual component failures in $\mathcal{C}'$  produce rough errors (see Fig.~\ref{fig:rough-smooth}).  To get the actual value of $p^{(2)}$, we need to add in the right-hand side (RHS) of~(\ref{eq:p_success_ii}), the probabilities of the events for all the subsets $\mathcal{C}'  \subseteq \mathcal{C}_{i \to j}, |\mathcal{C}'| \geq 2$, where each component in $\mathcal{C}'$ produces a rough error individually but the total error is a smooth error. As the probability of such an event is upper bounded by $\prod_{C \in \mathcal{C}'} p_C$, we expect the RHS in~(\ref{eq:p_success_ii}) to be a good lower bound of $p^{(2)}$. 
\end{remark}

\smallskip In the next paragraph, we give an approximation of $p^{(1)}$.
\paragraph{Rough error probability due to previous recursion levels.}

We first estimate the errors that remain on the input of $B_{i \to j}$ due to the smooth errors in previous recursion levels, \emph{i.e.}, $B_{0 \to i}$. We then estimate the probability that this remaining error is a rough error with respect to $B_{i \to j}$ (hence, giving an estimate of $p^{(1)}$).


\smallskip To estimate the remaining error on the input of $B_{i \to j}$,  we simply concatenate the smooth error channels corresponding to components in $\mathcal{C}_{0 \to i}$. The resulting channel after concatenation is a Pauli channel as given in Lemma~\ref{lem:smooth_total_channel}, which is proven in Appendix~\ref{app:proof-lemmas}.
%
%

\begin{lemma} \label{lem:smooth_total_channel}
Let $\mathcal{W}_{0 \to i}: L(H_{\mathcal{S}_{0 \to i}}) \to L(H_{\mathcal{S}_{0 \to i}})$ be the quantum channel that corresponds to the concatenation of smooth error channels for all $C \in \mathcal{C}_{0 \to i}$, i.e., $\mathcal{W}_{0 \to i} =  \mathcal{W}_{C_1} \circ \dots \circ \mathcal{W}_{C_{|\mathcal{C}_{0 \to i}|}}$. Then, $\mathcal{W}_{0 \to i} = \otimes_{q \in \mathcal{S}_{0 \to i}} \mathcal{W}_q$ (i.e., $\mathcal{W}_{0 \to i}$ acts independently and identically on qubits in $\mathcal{S}_{0 \to i}$), where $\mathcal{W}_q$ is a Pauli channel, whose $X$, $Y$ and $Z$ error probabilities are, respectively, upper bounded by $p^{0 \to i}_x, p^{0 \to i}_y$ and $p^{0 \to i}_z$, which are as follows,
\begin{align}
p^{0 \to i}_x &= \begin{cases}
1- (1-p)(1 - 2p/15)^{i} , & \text{if $\sum_{t=1}^i b_t = 0$.} \\
1-(1 - 2p/15)^{(i - k^{\min}_{0 \to i})+1}, & \text{if $\sum_{t=1}^i b_t \neq 0$, and $b_i = 0$.}  \\
2p/15, & \text{if $b_i = 1$.} 
\end{cases} \label{eq:prep_err_x}\\
p^{0 \to i}_y &= 2p/15. \label{eq:prep_err_y}  \\
p^{0 \to i}_z &= \begin{cases}
2p/15, & \text{if  $b_i = 0$.} \\
1-(1 - 2p/15)^{(i - k^{\min}_{0 \to i})+1}, & \text{if $b_i = 1$.} 
\end{cases} \label{eq:prep_err_z} 
\end{align}
\end{lemma}

\smallskip As explained in Remark~\ref{rem:rough-smooth-err} and Fig.~\ref{fig:rough-smooth}, a set of component failures may produce a smooth error even though individual component failures produce rough errors. Such smooth errors are not taken into account in Lemma~\ref{lem:smooth_total_channel}, therefore, the channel $\mathcal{W}_{0 \to i}$ therein underestimates the actual errors on the output of $B_{0 \to i}$. However, as the probability of such an event is exponentially small in the number of component failures, we expect $\mathcal{W}_{0 \to i}$ approximates well the actual channel acting on the output of $B_{0 \to i}$.



We now estimate the probability that the channel $\mathcal{W}_{0 \to i}$ in Lemma~\ref{lem:smooth_total_channel} produces a rough error with respect to $B_{i \to j}$. Note that an $X$ (or $Z$) error on the qubit $q \in \mathcal{S}_{i \to j}$ at the input of $B_{i \to j}$, will flip the measurement outcome of the next Pauli $Z \otimes Z$ (or $X \otimes X$) measurement. Therefore, using~(\ref{eq:prep_err_x}),~(\ref{eq:prep_err_y}), and~(\ref{eq:prep_err_z}), the probability of preexisting error on a qubit $q$ to be rough in $B_{i \to j}$ is upper bounded as,
\begin{equation} \label{eq:p_D_pre}
p_{\text{pre}} = \begin{cases}
p^{0 \to i}_y + p^{0 \to i}_z, & \text{ if $\sum_{t={i+1}}^j b_t = 0$.} \\
p^{0 \to i}_x + p^{0 \to i}_y, & \text{ if $\sum_{t={i+1}}^j b_t = j-i$.} \\
p^{0 \to i}_x + p^{0 \to i}_y + p^{0 \to i}_z, & \text{otherwise}.
\end{cases}
\end{equation}
%
%
Using~(\ref{eq:p_D_pre}) and the fact that there are $2^j$ qubits in $\mathcal{S}_{i \to j}$, the probability that the preexisting errors  do not produce a rough error with respect to $B_{i \to j}$ is given by,
\begin{equation} \label{eq:p_success_i}
p^{(1)} \approx (1-p_{\text{pre}})^{2^j}.
\end{equation}
Note that~(\ref{eq:p_success_i}) is an approximation due to the fact that several individual rough errors may be one smooth error when considered together, as previously explained.

\medskip Finally, from~(\ref{eq:p_success_i}) and~(\ref{eq:p_success_ii}), the probability that no-rough error happens in $B_{i \to j}$ (in other words, the success probability of $B_{i \to j}$ using Assumption~\ref{ass:err_det}) can be approximated as follows,
\begin{equation} \label{eq:p_success}
p^s_{i \to j} \approx (1-p_{\text{pre}})^{2^j} \prod_{C \in \mathcal{C}_{i \to j}}(1-p_C).
\end{equation}

\subsection{Preparation rate of the factory preparation} \label{sec:fact-prep-rate}
Using~(\ref{eq:p_success}), we now estimate the preparation rate for the asymptotically large factory size, that is, $T \to \infty$ as in~(\ref{eq:fact_prep_T}). Suppose after the $i_k^{th}, i_k \in n_{sch}$ scheduling recursion level, we have total $T_{i_k}$ successfully prepared $\pone$ code-states of length $2^{i_k}$. As in Procedure~\ref{prot:fact_prep}, we split successfully prepared code-states into $[T_{i_k}/(2^{i_{k+1} -i_k})]$ groups, each group containing $2^{i_{k+1} - i_k}$ code-states. We then apply $B_{i_k \to i_{k+1}}$ on each group to prepare $\pone$ code-states of length $2^{i_{k+1}}$. Using~(\ref{eq:p_success}) and the law of large numbers, we may estimate the number of prepared state after $B_{i_k \to i_{k+1}}$ as follows, 
\begin{equation} 
T_{i_{k + 1}} \approx [T_{i_k}/(2^{i_{k+1} - i_k})] p^s_{i_k \to i_{k+1}}. \label{eq:num_succ}
\end{equation}
Using~(\ref{eq:num_succ}), we have,
\begin{equation} \label{eq:num_succ1}
p^s_{i_k \to i_{k+1}} \approx \frac{T_{i_{k + 1}}  (2^{i_{k+1} - i_k})}{T_{i_k}}. 
\end{equation}
Further, applying~(\ref{eq:num_succ1}) recursively, we may get an estimate of the preparation rate in~(\ref{eq:fact_prep_T}), as follows (we take $i_0 = 0$ and recall $i_{|n_{sch}|} = n$),
\begin{align}
 p^{n_{sch}}_{\text{fact}} & = \frac{T_{i_{|n_{sch}|}}}{T} \\
& = \displaystyle\prod_{k = 0}^{i_{|n_{sch}| - 1}}  \frac{T_{i_{k+1}} (2^{i_{k+1} - i_k})}{T_{i_k}} \\
& \approx \displaystyle\prod_{k = 0}^{i_{|n_{sch}|} - 1} p^s_{i_k \to i_{k+1}}, \label{eq:ratio_int}
\end{align}
where we have used $T_{i_0} = T_0 = NT$ in the second equality, and approximation in the last line follows from~(\ref{eq:num_succ1}). Therefore, from~(\ref{eq:ratio_int}), the factory preparation rate for $T \to \infty$ is simply multiplication of the success probabilities of the blocks $B_{i_k \to i_{k+1}}$, $ k = 0, \dots, i_{|n_{sch}|}-1$. Further, from~(\ref{eq:p_success}) and~(\ref{eq:ratio_int}), we may roughly see as follows why factory preparation may improve the preparation rate compared to preparation in~\cite{goswami2022fault}. As noted before, preparation in~\cite{goswami2022fault} corresponds to $n_{sch} = \{0, n\}$, for which we get $  p^{n_{sch}}_{\text{fact}} \approx  p^s_{0 \to n}$ from~(\ref{eq:ratio_int}). Further from~(\ref{eq:p_success}), for $p^s_{0 \to n}$, we need to consider components in the set $\mathcal{C}_{0 \to n}$ for error detection, while for $\prod_{k = 0}^{i_{|n_{sch}|} - 1} p^s_{i_k \to i_{k+1}}$ corresponding to a $n_{sch} = \{ i_0, \dots, i_k, \dots, i_{|n_{sch}|} \} \supset \{0, n\}$, we need to consider components in the smaller set $\mathcal{C}_{0 \to i_1} \cup \cdots \cup \mathcal{C}_{i_{k-1} \to i_k} \cup \mathcal{C}_{i_k \to i_{k+1}} \dots \cup \mathcal{C}_{i_{|n_{sch}|-1} \to i_{|n_{sch}|}} \subset \mathcal{C}_{0 \to n}$.


\subsection{Preparation error probabilities of the factory preparation}
The remaining error on the final prepared state is due to the smooth errors in $B_{0 \to n}$, hence given by the quantum channel $\mathcal{W}_{0 \to n}$ as in Lemma~\ref{lem:smooth_total_channel}. The action of $\mathcal{W}_{0 \to n}$ on each qubit $q \in \mathcal{S}_{0 \to j}$ corresponds to the Pauli channel, with the $X, Y$, and $Z$ error probabilities $p_x^{0 \to n}, p_y^{0 \to n}$, and  $p_x^{0 \to n}$ according to~(\ref{eq:prep_err_x}),~(\ref{eq:prep_err_y}),
and~(\ref{eq:prep_err_z}), respectively. Therefore, the preparation $X$ and $Z$ error probabilities in~(\ref{eq:avg_X}) and~(\ref{eq:avg_Z}), respectively, are given by,
\begin{align}
p_X^{prep} &= p_x^{0 \to n} + p_y^{0 \to n}. \label{eq:prep_x_erate}\\
p_Z^{prep} &= p_y^{0 \to n} + p_z^{0 \to n}. \label{eq:prep_z_erate}
\end{align}

\section{Numerical results} \label{sec:num-res}

In this section, we present our numerical results regarding the factory preparation rate and logical error rates, using our theoretical estimates in Section~\ref{sec:estimates_theory} as well as a Monte-Carlo simulation, considering the circuit level depolarizing noise model from Section~\ref{sec:prep-n}.
\subsection{Preparation rate}

\begin{figure}[!t]
 \begin{subfigure}[b]{0.49\textwidth}
      \includegraphics[width=\textwidth]{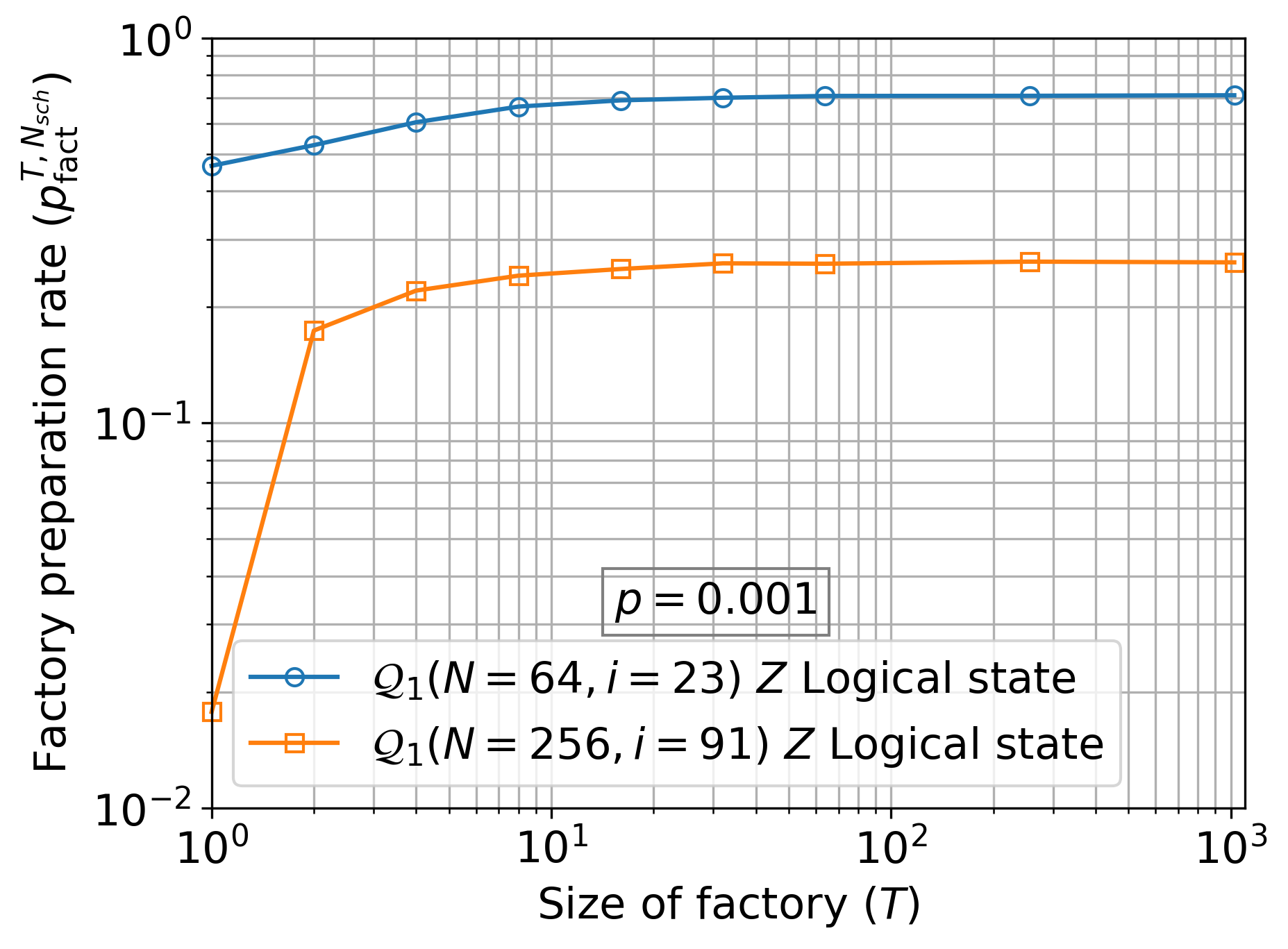}
      \captionsetup{justification=centering}
      \caption{}
      \label{fig:fact_rate_num}
    \end{subfigure}
 \begin{subfigure}[b]{0.50\textwidth}
      \includegraphics[width=\textwidth]{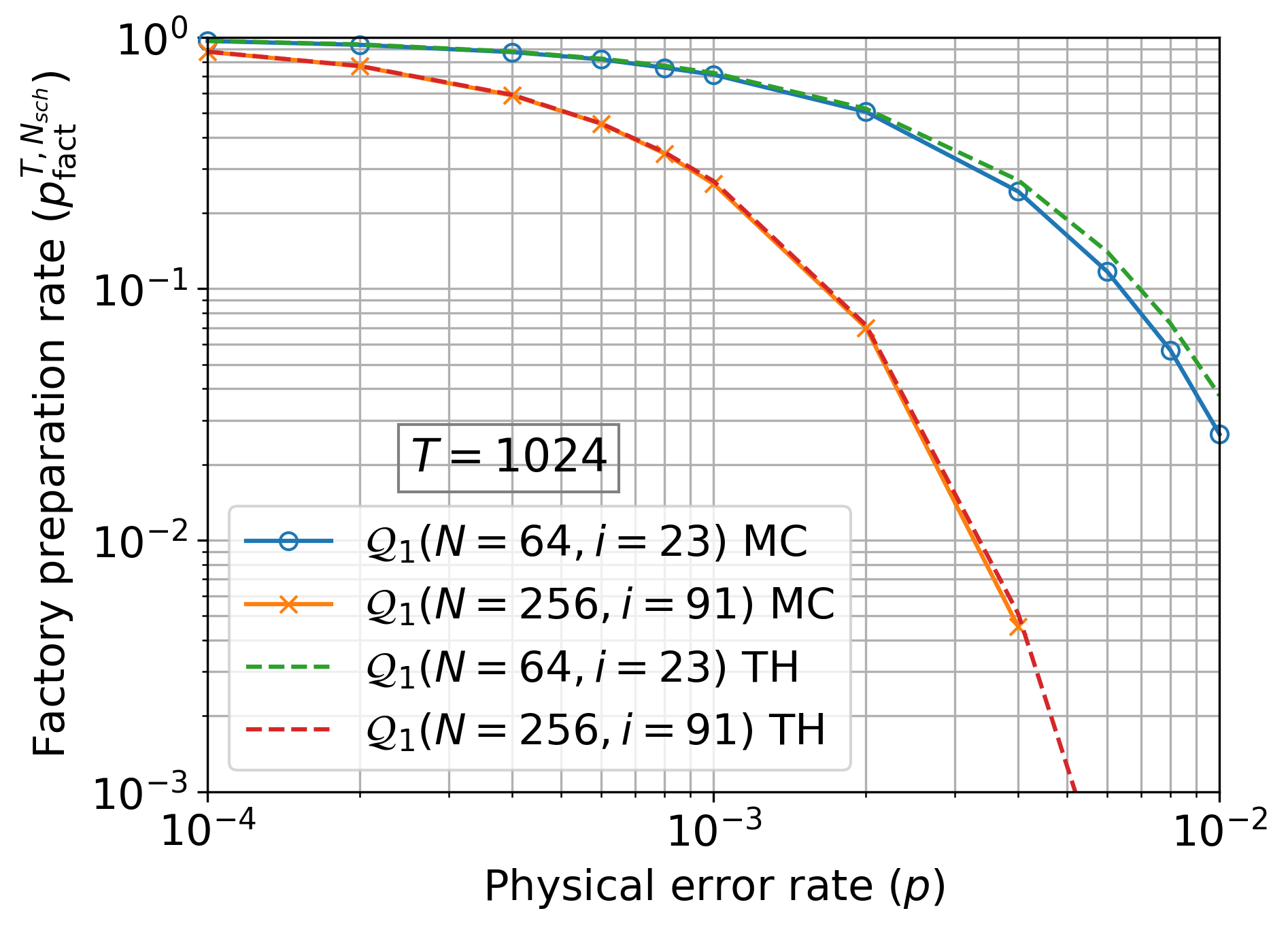}
      \captionsetup{justification=centering}
      \caption{}
       \label{fig:fact_rate_saturated_num}
    \end{subfigure}
\caption{Factory preparation rate: we consider codes $\pone(N = 64, i = 23)$  and $\pone(N = 256, i = 91)$, and scheduling sets $n_{\text{sch}} = \{2, 4, 6\}$ and $n_{\text{sch}} = \{2, 4, 6, 8\}$, respectively, for the factory preparation according to Procedure~\ref{prot:fact_prep}. Figure~(a) shows factory preparation rate with respect to the size of the factory $T$ for a fixed value of physical error rate $p = 10^{-3}$, obtained using a Monte-Carlo simulation. Figure~(b) shows factory preparation rate with respect to the physical error rate $p$ for a fixed value of the factory size $T = 1024$, obtained using a Monte-Carlo simulation and also using theoretical estimates from Section~\ref{sec:fact-prep-rate}.}
\label{fig:prep_rate}
\end{figure}

For the Monte-Carlo simulation of the preparation rate, we proceed as follows. We simulate $R$ times the factory preparation according to Procedure~\ref{prot:fact_prep}, for a factory size $T$. Let $t_i$ be the number of successfully prepared states for the $i^{th}$ instance of the factory preparation, $1 \leq i \leq R$. Then, we determine $p^{T, n_{\text{sch}}}_{\text{fact}}$ as follows,
\begin{equation} \label{eq:fact_prep-1}
p^{T, n_{\text{sch}}}_{\text{fact}} = \frac{1}{RT} \sum_{i = 1}^R t_i.
\end{equation}
We obtain the value of $p^{T, n_{\text{sch}}}_{\text{fact}}$ for $T$ values $1, 2^1, \dots, 2^{10}$. For the fixed physical error rate $p = 10^{-3}$, the factory preparation rate $p^{T, n_{\text{sch}}}_{\text{fact}}$, with respect to $T$, is shown in Fig.~\ref{fig:fact_rate_num} for $\pone(N = 64, i = 23)$ and $\pone(N = 256, i = 91)$, with scheduling sets $n_{\text{sch}} = \{ 2, 4, 6\}$ and $n_{\text{sch}} = \{2, 4, 6, 8\}$, respectively. The information positions are chosen according to~\cite[Table I]{goswami2022fault} for ignoring correlations.

\smallskip We observe that $p^{T, n_{\text{sch}}}_{\text{fact}}$ increases with respect to $T$ in the beginning, and then it saturates. The saturated value corresponds to the preparation rate with respect to $T \to \infty$, \emph{i.e.}, $p^{n_{\text{sch}}}_{\text{fact}}$ from~(\ref{eq:fact_prep_T}). We take the saturated value to be the value of $p^{T, n_{\text{sch}}}_{\text{fact}}$ for $T = 2^{10}$. 

\smallskip The difference between the saturated value and the value of $p_{\text{prep}}$ from~(\ref{eq:prep_rate}) ($i.e.$, $p^{T, n_{\text{sch}}}_{\text{fact}}$ for $T = 1$) is quite significant, especially for $N = 256$. In particular, for $N = 64, 256$, the value of $p_{\text{prep}}$ is around $47\%, 2\%$, respectively, while the saturated value of  $p_{\text{prep}}$ is around $70\%, 27\%$. Therefore, the factory preparation provides significant improvement in the preparation rate compared to~\cite[Procedure 2]{goswami2022fault}. Further, the saturation happens rather quickly, for example the value of $p^{T, n_{\text{sch}}}_{\text{fact}}$ for $T = 8$ is already quite close to its value for $T = 1024$ for both $N = 64, 256$.  This means that we do not need a large factory size to get the increased preparation rate of the factory preparation\footnote{Note that the inverse of the preparation rate contributes to the qubit overhead of the preparation, hence, the factory preparation reduces significantly the qubit overhead of $\pone$ code-state preparation. In particular for $N = 256, p = 10^{-3}$, it reduces the qubit overhead by a factor of around $13$. }. 

\smallskip We further obtain the value of $p^{n_{\text{sch}}}_{\text{fact}}$ using our theoretical estimate in~(\ref{eq:p_success}) and~(\ref{eq:ratio_int}). In Fig.~\ref{fig:fact_rate_saturated_num}, we have presented both the Monte-Carlo and theoretical values of $p^{n_{\text{sch}}}_{\text{fact}}$ with respect to the physical error rate $p$. We observe that the curves corresponding to Monte-Carlo simulation and theoretical estimates are very close,  thus validating that our method of theoretical estimation is a good approximation of reality.

\subsection{Logical error rate}
%
In this section, we estimate the logical error rates of $\pone$ codes, using the Steane error correction, which uses the ancilla code-states (logical $\ket{0}$ and $\ket{+}$ states), prepared by the factory preparation according to Procedure~\ref{prot:fact_prep}. 

\begin{figure}[!t]
\,\hfill\includegraphics[width = 0.8\linewidth]{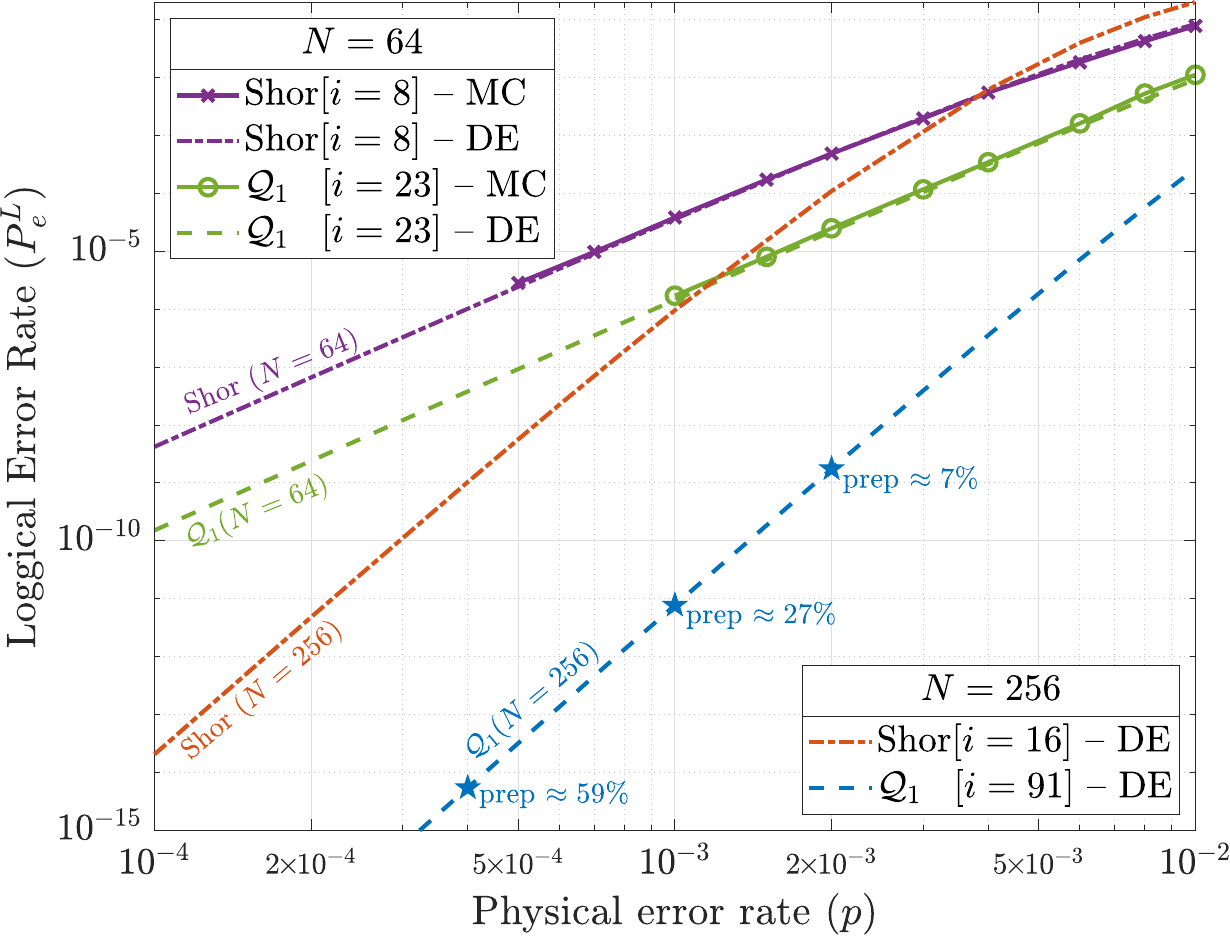} \hfill\,
\caption{Logical error rates of codes $\pone(N = 64, i = 23)$ and $\pone(N = 256, i = 91)$, using Steane error correction, where ancilla  code-states are prepared using the factory preparation in Procedure~\ref{prot:fact_prep}.}
\label{fig:log_rate_num}
\end{figure}

To do so, we use the density evolution technique as in~\cite[Section V.D]{goswami2022fault}. Precisely, using the $X$ and $Z$ error probabilities on the prepared states, given by~(\ref{eq:prep_x_erate}) and~(\ref{eq:prep_z_erate}), we first estimate the input error probability for the two decoders used within the Steane error correction procedure, as described in~\cite[Section V.C and Section V.D]{goswami2022fault} (see Eqs. (105) and (106) therein). Then, we use density evolution to estimate the output error probability of the two decoders, from which we determine the $X$ and $Z$ logical error rates, $P^L_X$ and $P^L_Z$~\cite[Eq. (107)]{goswami2022fault}, and then the (total) logical error rate $P^L_e = P^L_X + P^L_2 - P^L_X P^L_Z$.

\begin{table}[!t]
\begin{center}
\captionsetup{justification=centering}
\caption{Preparation and logical error rates for $N = 1024$.} 
\label{table:N=1024}
\begin{tabular}{|c|c|c|c|c|}
\hline
$p$ & $0.001$ & $0.0004$ & $0.0002$ & $0.0001$ \\
\hline
$p_{\text{fact}}^{n_{sch}}$ & $0.5\%$ & $12\%$ & $35\%$ & $59\%$ \\
\hline
$P_e^L$ & $4.08 \times 10^{-22}$ & $2.04 \times 10^{-28}$ & $3.23 \times 10^{-33}$ & $2.69 \times 10^{-38}$ \\
\hline
\end{tabular}
\end{center} 
\end{table}


In Fig.~\ref{fig:log_rate_num}, we present the logical error rate vs. physical error rate curves for $N= 64, 256$. For $N = 64$, we have included the density evolution based curves as well as the Monte-Carlo simulation based curves from~\cite[Fig. 4]{goswami2022fault}. We observe that the density evolution curves virtually superimpose  the Monte-Carlo curves, therefore substantiating our theoretical method of estimating preparation error probabilities.

\smallskip For $N = 256$, we have only included the density evolution curves, as the logical error rates are very small to be simulated using the Monte-Carlo simulation, and also the preparation rates are comparatively smaller. As expected, the $\pone$ codes perform much better than the Shor-$\pone$ codes. Note that the performance of the $\pone$ code for $N = 64$ is better than the performance of the  Shor-$\pone$ code for $N = 256$, down to a physical error rate $p = 10^{-3}$. Remarkably, the $\pone$ code for $N = 256$ achieves logical error rates around $10^{-11}$ and $10^{-15}$ for physical error rates $10^{-3}$ and $3 \times 10^{-4}$, respectively, which is very promising for the practical large-scale fault-tolerant quantum computation~\cite{fowler2012surface}. 

\smallskip Finally, for $N = 1024$, our numerical results are given in Table~\ref{table:N=1024}. For a practically interesting range of physical error rates $p \in [10^{-4}, 10^{-3}]$, we observe that $\pone$ code of length $N = 256$ is the best choice due to good preparation rate and sufficiently low logical error rates. Increasing the code-length may be useful for physical error rates above $10^{-3}$, however in this case the $N=1024$ code is penalized by its poor preparation rate.   

\subsection{Comparison with the surface code}
 
\begin{figure}[!b]
\,\hfill\includegraphics[width = 0.8\linewidth]{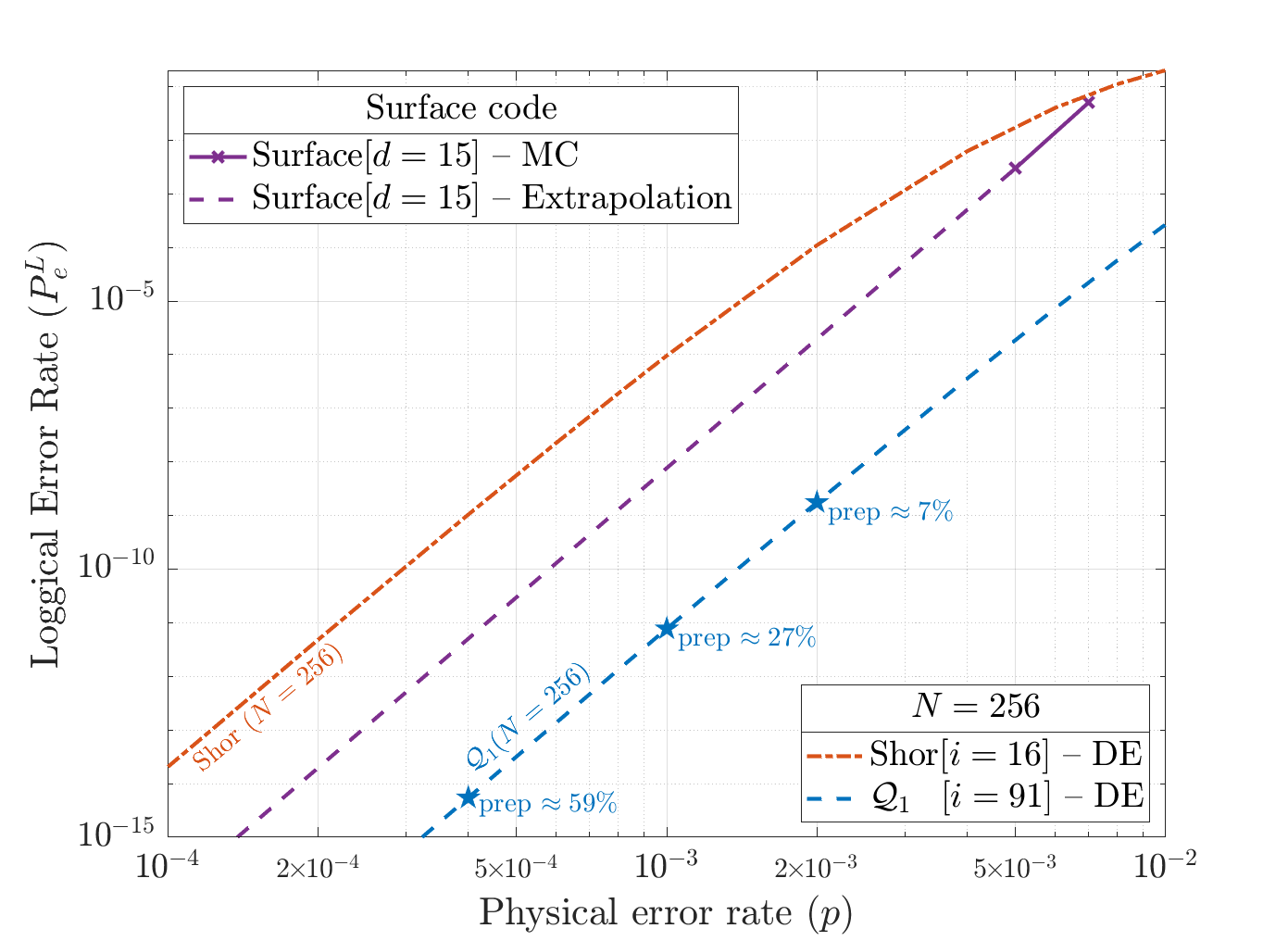} \hfill\,
\caption{Comparison of logical error rates of $\pone(N = 256, i = 16)$ and $\pone(N = 256, i = 91)$, with that of the surface code, with code-length $N= 225$, and minimum distance $d=15$ from~\cite{fowler2012surface}.}
\label{fig:log_rate_comp-sur}
\end{figure}

In Fig.~\ref{fig:log_rate_comp-sur}, we compare the error correction performance of $\pone$ and Shor $\pone$ codes of length $N = 256$ and minimum distance $d = 16$, with that of a surface code with minimum distance $d=15$, assuming a circuit level depolarizing noise model.  For the surface code, the simulation results are taken from~\cite{fowler2012surface}. Note that taking $d=16$ for the surface code only increases the code-length $N$, but not the error correction performance. For $d=15$, the code-length reported in~\cite{fowler2012surface} is $N=421$, however, it can be reduced to  $N=225$, by considering a rotated variant of the surface code. The logical error rate of the surface code is simulated in~\cite{fowler2012surface} down to a physical error rate $p = 5 \times 10^{-3}$. We have extrapolated the logical error rate for lower physical error rates using $P_e^L = c (p / p_{th}) ^{\frac{d+1}{2}}$, as proposed in~\cite{fowler2012surface}, where   $p_{th} = 0.0057$ is the reported surface code threshold.

\smallskip We can see that the $\pone$ code outperforms the surface code by about three orders of magnitude. For example, for practically interesting physical error rates $p = 10^{-3}$ and $p = 5 \times 10^{-4}$, the corresponding logical error rates for the surface code are around $10^{-8}$ and $5 \times 10^{-11}$, while for $\pone$ codes are around $10^{-11}$ and $5 \times 10^{-14}$.  This is an encouraging result for $\pone$ codes and shows that they are of independent interest in the context of fault tolerant quantum computing.  

However, we should consider the above comparison carefully, as error correction procedures are different for $\pone$ and surface codes. While Steane error correction is natural for $\pone$ codes, generator measurement-based error correction is natural for surface codes due to their small weight generators. On the one hand, Steane error correction is advantageous in the sense that one round of syndrome extraction is enough for error correction, while for generator measurement-based error correction, several rounds of syndrome extraction are needed. Therefore, error-correction for polar codes can be faster than for surface codes.  On the other hand, generator measurement-based error correction is advantageous in the sense that ancilla qubits can be intercalated between data qubits and are directly reusable after each round of syndrome extraction. However, Steane error correction needs a separate ancilla factory running, to produce ancilla states needed for error correction. Ancilla qubits used in a round of error correction will be  moved back to the factory so that they are reused. 


\smallskip Finally, we note that the factory preparation (see also Fig.~\ref{fig:error_det_big}) requires distant $\cnot$ gates, \emph{i.e.} interaction between non-neighboring qubits, as opposed to the surface codes.  Although distant $\cnot$ gate is possible on some potential quantum systems such as ion-traps~\cite{ryan2021realization,postler2022demonstration}, there are  ways to circumvent this for quantum systems with local interaction constraint such as by applying swap gates~\cite{sigillito2019coherent, asai2023device} or physically moving qubits around~\cite{jadot2021distant, seidler2022conveyor, boter2022spiderweb, bluvstein2022quantum,bluvstein2023logical}. 

\smallskip Recently, quantum error correcting codes such as small surface and color codes, as well as three-dimensional codes,  have been implemented on a reconfigurable quantum architecture (based on Rydberg atoms), with storage, entangling and readout zones~\cite{bluvstein2022quantum,bluvstein2023logical}.  Qubits therein are moved around within a zone or between zones to achieve long-range connectivity. Moreover, it is worth noticing that the $3$-dimensional $[[8,3,2]]$ code implemented in~\cite{bluvstein2023logical} is actually very similar to a length-8 polar code, both codes using hypercube connectivity. As $\pone$ codes provide better error correction performance exploiting distant operations, they are naturally suited to this kind of architecture. Moreover, the reconfigurable quantum architecture allows to move blocks of qubits in parallel, realizing transversal $\cnot$ gate on two code blocks in parallel. As our preparation is based on recursively applying transversal measurements between two blocks of qubits, it may be implemented in a similar way.

\section{Discussion} \label{sec:conc}

We have proposed a factory preparation of $\pone$ code-states, which is shown to be an useful extension of the measurement based preparation in~\cite{goswami2022fault}, providing much better preparation rates comparatively. Its better preparation rate owes to a scheduling step, which makes clever use of the $\pone$ code-states prepared at the intermediate levels of recursion. Thanks to the factory preparation, we are able to prepare code-states of lengths $N = 256, 1024$, with reasonably high preparation rates for a practically interesting physical error rate range  $10^{-4}-10^{-3}$. Further, it is shown that for $N = 256, 1024$ and a physical error rate in the range $10^{-4}-10^{-3}$, the $\pone$ code achieves a logical error rate below $10^{-15}$, which is currently estimated to be the required logical error rate for large-scale fault-tolerant quantum computation~\cite{fowler2012surface}. Due to a higher preparation rate, smaller number of qubits and also achieving a sufficiently low logical error rate, the code-length $N=256$ is the best $\pone$ code in this physical error range. 

To estimate the preparation rate and logical error rates of $\pone$ codes, we have used a theoretical framework based on the new notions of smooth and rough errors. It is shown that estimates based on our theoretical framework fit well the estimates obtained using Monte-Carlo simulations, therefore substantiating the accuracy of the theoretical framework. We note that our notions of smooth and rough errors are not particular to $\pone$ codes and therefore, as a natural future direction, it would be interesting to analyze other fault-tolerant protocols using these notions, especially the ones based on error detection~\cite{knill2005scalable,chao2018quantum}. 
%
%
\section{Acknowledgement}
This work was supported by the QuantERA grant EQUIP, by the French Agence Nationale de la Recherche, ANR-22-QUA2-0005-01.

\appendix

\section{Proofs of Lemmas~\ref{lem:int-eff-err},~\ref{lem:p-w-int-m},~\ref{lem:p-w-cnot} and~\ref{lem:smooth_total_channel}} \label{app:proof-lemmas}

\begin{figure}[!b]
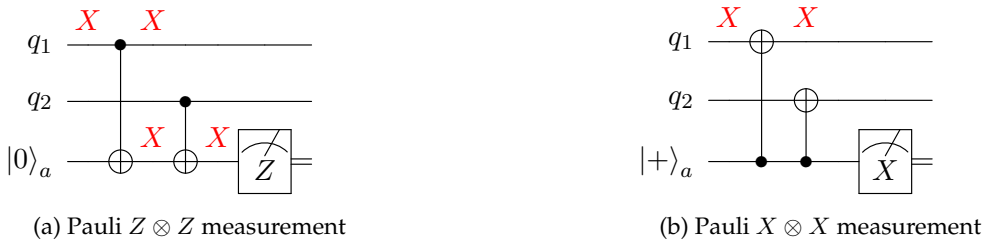

\centering
\begin{subfigure}[b]{0.5 \linewidth}
\centering
\,\hfill\input{qcircuits/circuit_lemma_initialization}\hfill \,
\captionsetup{justification=centering}
\caption{Pauli $Z \otimes Z$ measurement}
\label{fig:proof_lemma_initializationZZ}
\end{subfigure}\hfill%
\begin{subfigure}[b]{0.5 \linewidth}
\centering
\,\hfill\input{qcircuits/circuit_lemma_initializationXX}\hfill \,
\captionsetup{justification=centering}
\caption{Pauli $X \otimes X$ measurement}
\label{fig:proof_lemma_initializationXX}
\end{subfigure}
\captionsetup{justification=centering}
\caption{Propagation of initialization errors through Pauli $Z \otimes Z$ and $X \otimes X$ measurement.}
\label{fig:proof_lemma_initialization}
\end{figure}

\paragraph{Proof of Lemma~\ref{lem:int-eff-err}:} Lemma~\ref{lem:int-eff-err} follows from the fact that the initialization at the zeroth level of recursion produces an $X$ error, which flips the outcome of a Pauli $Z \otimes Z$, while it does not have any effect on Pauli $X \otimes X$ measurement as shown in Fig.~\ref{fig:proof_lemma_initialization}.

\paragraph*{Proof of Lemma~\ref{lem:p-w-int-m}:}
Consider the circuit implementing Pauli $Z \otimes Z$ measurement from Fig.~\ref{fig:mZZ_mXX_notation_and_circuit}. Note that a failure in initialization on the ancilla qubit produces an $X$ error on it, which will flip the outcome of the next Pauli $Z$ measurement. A failure in Pauli $Z$ measurement produces an $X$ error just before the measurement, hence it will also flip the measurement outcome. Therefore, the measurement outcome will be flipped with probability $1$ if an initialization or an measurement error happens on the ancilla qubit. Therefore, it follows that,
\begin{align}
p_C &= p. \label{eq:p-det-M}  \\ 
\mathcal{W}_C &= I. \label{eq:W-M}
\end{align} 
Similarly, it can be proven for Pauli $X \otimes X$ measurement.


\paragraph*{Proof of Lemma~\ref{lem:p-w-cnot}:}
We  prove below Lemma~\ref{lem:p-w-cnot} for the $\cnot$ gate applied at the time step $t = 2$ of a recursion level in $B_{i \to j}$, with Pauli  $Z \otimes Z$ measurements. It can be similarly seen that Lemma~\ref{lem:p-w-cnot} also holds for $t = 3$, and as well as for the $\cnot$ gates applied in a Pauli $X \otimes X$ measurement.

\begin{figure}[!t]
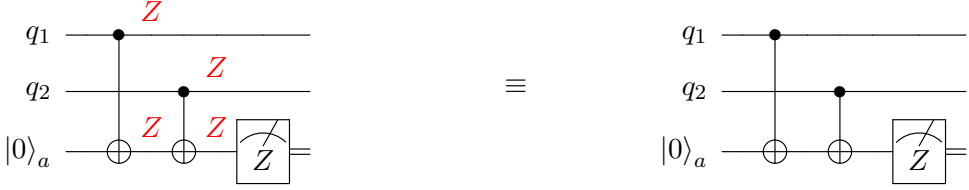
 
\centering
\begin{subfigure}[b]{0.48\linewidth}
\,\hfill\input{qcircuits/propagation_err_ZZ}\hfill\,
\end{subfigure}\hfill%
\raisebox{-2.0em}{$\equiv$}\hfill%
\begin{subfigure}[b]{0.48\linewidth}
\,\hfill\input{qcircuits/propagation_err_YZ}\hfill\,
\end{subfigure}
\captionsetup{justification=centering}
\caption{A fault in $\cnot_{q_1 \to a}$ leading to a stabilizer error.}
\label{fig:proof_lemma_cnot}
\end{figure}

Consider a Pauli $Z \otimes Z$ measurement applied at some $k^{th}$ recursion level corresponding to $B_{i \to j}$. We shall denote the data qubits on which Pauli $Z \otimes Z$ measurement acts on by $q_1, q_2$ and the ancilla qubit by $a$, as in Fig.~\ref{fig:proof_lemma_cnot}. Below, we  classify the errors produced by the first $\cnot$ gate $\cnot_{q_1 \to a}$ into rough and smooth errors, as per Definition~\ref{def:eff_err}, and then, using the set of rough and smooth errors, we compute $p_C$ and $\mathcal{W}_C$ for $C = \cnot_{q_1 \to a}$.

\smallskip Firstly, note that if a failure in $C$ produces the error $Z_{q_1} \otimes Z_{a}$, it propagates as $Z_{q_1} \otimes Z_{q_2}$ after the measurement, which is a stabilizer of the quantum state corresponding to the joint system $q_1q_2$, and hence this error can be ignored (see Fig.~\ref{fig:proof_lemma_cnot}). Further, if a failure in $C$ produces an $X$ or $Y$ error on the ancilla qubit $a$, it will flip the measurement outcome of the corresponding Pauli $Z \otimes Z$ measurement, hence such an error is a rough error. Precisely, the following errors produce an $X$ or $Y$ error on the ancilla,
\begin{equation}
I_{q_1} \otimes X_{a}, X_{q_1} \otimes X_{a}, Z_{q_1} \otimes X_{a}, Y_{q_1} \otimes X_{a}, I_{q_1} \otimes Y_{a}, X_{q_1} \otimes Y_{a}, Z_{q_1} \otimes Y_{a}, Y_{q_1} \otimes Y_{a}
\end{equation}

The remaining errors, \emph{i.e.}, $ X_{q_1} \otimes I_{a}, Y_{q_1} \otimes I_{a}, Z_{q_1} \otimes I_{a}, I_{q_1} \otimes Z_{a}, X_{q_1} \otimes Z_{a}, Y_{q_1} \otimes Z_{a}$ do not flip the outcome of the measurement and propagate to the following errors  after the measurement (up to the stabilizer $Z_{q_1} \otimes Z_{q_2}$, similar to Fig.~\ref{fig:proof_lemma_cnot}),  $X_{q_1} \otimes I_{q_2}, Y_{q_1} \otimes I_{q_2}, Z_{q_1} \otimes I_{q_2}, Z_{q_1} \otimes I_{q_2}, Y_{q_1} \otimes I_{q_2}, X_{q_1} \otimes I_{q_2}$, respectively. 
Hence, the remaining errors only act non-trivially on the qubit $q_1$, and they  correspond to a depolarizing channel, with error probability $6p/15$ as follows,   
$\mathcal{D}_{q_1}^{(6p/15)}(\rho) := (1 - \frac{6p}{15}) \rho + \frac{2p}{15} (X \rho X + Y \rho Y + Z \rho Z)$.
Some or all of the remaining errors may also be rough errors depending on whether they flip a measurement outcome at one of the next recursion levels. We have the following four cases in order.
\paragraph*{If $k$ is the last recursion level, \emph{i.e.}, $k = j$.}
In this case, $p_C$ is simply given by the errors that flip the measurement outcome of the corresponding Pauli $Z \otimes Z$ measurement at the $k^{th}$ recursion level. Therefore $p_C = 8p/15$. Further, $\mathcal{W}_C$ is the remaining noise channel after the $k^{th}$ recursion level. Therefore, $\mathcal{W}_C = I_{\mathcal{S}_{i \to j}\setminus {q_1}} \otimes \mathcal{D}_{q_1}^{(6p/15)}$.

\paragraph*{If $k^{\min}_{i \to j} \leq k <  j$ with $b_j = 0$ (hence, only Pauli $X \otimes X$ measurements after the $k^\textbf{th}$ recursion level.)}

In this case, the remaining $Z$ error after the $k^{th}$ level of recursion will flip a Pauli $X \otimes X$ measurement at the $(k+1)^{th}$ level of recursion, while the remaining $X$ errors will not flip any measurement outcome. This implies that the depolarizing channel $\mathcal{D}_{q_1}^{(6p/15)}$ transforms into a bit-flip channel, denoted by $\mathcal{B}_{q_1}^{(2p/15)}$, acting on $q_1$ with error probability $2p/15$. Therefore, $p_C = 12p/15$ and $\mathcal{W}_C = I_{\mathcal{S}_{i \to j}\setminus {q_1}} \otimes \mathcal{B}_{q_1}^{(2p/15)}$.

\paragraph*{If $k^{\min}_{i \to j} \leq k <  j$ with $b_j = 1$ (hence, only Pauli $Z \otimes Z$ measurements after the $k^\textbf{th}$ recursion level.)}

In this case, the remaining $X$  error after the $k^{th}$ level of recursion, will flip a Pauli $Z \otimes Z$ measurement at the $(k+1)^{th}$ level of recursion, while the remaining $Z$ errors will not flip any measurement outcome. This implies that the depolarizing channel $\mathcal{D}_{q_1}^{(6p/15)}$ transforms into a phase-flip channel, denoted by $\mathcal{P}_{q_1}^{(2p/15)}$, acting on $q_1$ with error probability $2p/15$. Therefore, $p_C = 12p/15$ and $\mathcal{W}_C = I_{\mathcal{S}_{i \to j}\setminus {q_1}} \otimes \mathcal{P}_{q_1}^{(2p/15)}$.

\paragraph*{If $k < k^{\min}_{i \to j}$.}
In this case, we have both Pauli $Z \otimes Z$ and $X \otimes X$ measurements afterwards, hence both the remaining $X$ and $Z$ errors after the $k^{th}$ recursion level will be detected in one of the next recursion levels. Therefore, the depolarizing channel $\mathcal{D}_{q_1}^{(6p/15)}$ transforms into the identity channel. Hence, $p_C = 14p/15$ and $\mathcal{W}_C = I_{\mathcal{S}_{i \to j}\setminus {q_1}} \otimes I_{\mathcal{S}_{i \to j}}$.

%
%

In summary, we have the following for  $p^C_{D}$ and $\mathcal{W}_C$,
\begin{equation} \label{eq:eff_err_cnot_app}
 p^C_{D} = \begin{cases}
 8p/15, & \text{ if $k =  j$. }  \\
 4p/5, & \text{ if $  k^{\min}_{i \to j} \leq k <  j$.} \\
 14p/15, & \text{ otherwise.} \\
\end{cases}
\end{equation}
Further, $\mathcal{W}_C = I_{\mathcal{S}_{i \to j} \setminus {q}} \otimes \mathcal{W}_q$, where $\mathcal{W}_q$ is a quantum channel acting on $q$ as follows, 
\begin{equation} \label{eq:eff_noise_cnot_app}
\mathcal{W}_q = \begin{cases}
\mathcal{D}_{q}^{(6p/15)}, & \text{ if $k =  j$ }  \\
 \mathcal{B}_{q}^{(2p/15)}, & \text{ if $k^{\min}_{i \to j} \leq k <  j$, with $b_j = 0$.} \\
  \mathcal{P}_{q}^{(2p/15)}, & \text{if $k^{\min}_{i \to j} \leq k < j$, with $b_j = 1$.} \\
 I_{\mathcal{S}_{i \to j}}, & \text{ otherwise.} \\
\end{cases}
\end{equation}

%

\paragraph{Proof of Lemma~\ref{lem:smooth_total_channel}:} \smallskip From Lemma~\ref{lem:int-eff-err}, the smooth error channel corresponding to initialization at the zeroth level of recursion (with respect to $k^{\min}_{0 \to i}$) is the identity, except when $ \sum_{t = 1}^i b_t = 0$ (it implies that $k^{\min}_{0 \to i} = 1$), when it is a bit flip channel on the corresponding qubit at the output of $B_{0 \to i}$, as in~(\ref{eq:non_eff_int_zero}). Further, from Lemma~\ref{lem:p-w-int-m}, the smooth error channel corresponding to initialization and measurement of the ancilla qubit is always the identity. Furthermore, from Lemma~\ref{lem:p-w-cnot}, the smooth error channel corresponding to $\cnot$ gates for recursion levels $k < k^{\min}_{0 \to i}$ is also the identity. For $k \geq k^{\min}_{0 \to i}$, it is either a depolarizing, bit or phase-flip channel as given in~(\ref{eq:eff_noise_cnot}), acting on the data qubit on which the $\cnot$ gate acts. 
Hence, from~(\ref{eq:non_eff_int_zero}) and ~(\ref{eq:eff_noise_cnot}), and noting that one $\cnot$ gate acts on a data qubit $q \in \mathcal{S}_{i \to j}$ at each recursion level (see Fig.~\ref{fig:error_det_big}), it follows that the total smooth channel on each qubit $q \in \mathcal{S}_{0 \to i}$ is an (\emph{i.i.d.}) Pauli channel as follows,
\begin{equation}\label{eq:total_non_eff_channel}
\mathcal{W}_q = \begin{cases}
\mathcal{B}^{(p)}_q \circ (\mathcal{B}^{(2p/15)}_q \circ \stackrel{i}{\cdots} \circ \mathcal{B}^{(2p/15)}_q) \circ \mathcal{D}^{(6p/15)}_q, & \text{if $\sum_{t=1}^i b_t = 0$.}  \\
(\mathcal{B}^{(2p/15)}_q \circ \stackrel{i - k^{\min}_{0 \to i}}{\cdots} \circ \mathcal{B}^{(2p/15)}_q) \circ \mathcal{D}^{(6p/15)}_q, & \text{if $\sum_{t=1}^i b_t \neq 0$ and $b_i = 0$.}  \\
(\mathcal{P}^{(2p/15)}_q  \circ\stackrel{i - k^{\min}_{0 \to i}}{\cdots} \circ \mathcal{P}^{(2p/15)}_q) \circ \mathcal{D}^{(6p/15)}_q, & \text{if $b_i = 1$.}  
\end{cases}
\end{equation}
By simplifying~(\ref{eq:total_non_eff_channel}), we can see that $\mathcal{W}_q$ is a Pauli channel, whose  $X$, $Y$, and $Z$ error probabilities are, respectively, upper bounded by $p^{0 \to i}_x, p^{0 \to i}_y$, and $p^{0 \to i}_z$, which are given by,
\begin{align}
p^{0 \to i}_x &= \begin{cases}
1- (1-p)(1 - 2p/15)^{i} , & \text{if $\sum_{t=1}^i b_t = 0$.} \\
1-(1 - 2p/15)^{(i - k^{\min}_{0 \to i})+1}, & \text{if $\sum_{t=1}^i b_t \neq 0$ and $b_i = 0$.}  \\
2p/15, & \text{if $b_i = 1$.} 
\end{cases}\\
p^{0 \to i}_y &= 2p/15.  \\
p^{0 \to i}_z &= \begin{cases}
2p/15, & \text{if $b_i = 0$.} \\
1-(1 - 2p/15)^{(i - k^{\min}_{0 \to i})+1}, & \text{if $b_i = 1$.} 
\end{cases} 
\end{align}
We note that $p^{0 \to i}_y$ is equal to the $Y$ error probability. Furthermore, $p^{0 \to i}_x$ is equal to the $X$ error probability for $b_i = 1$, and $p^{0 \to i}_z$ is equal to the $Z$ error probability, when $b_i = 0$.
%
%
\printbibliography
\end{document}